\documentclass[aps,prc,reprint,superscriptaddress]{revtex4-2}

\usepackage{graphicx}%include figures
\usepackage{amsmath}
\usepackage{float}
\usepackage{paralist}
\usepackage{enumitem}
\usepackage{mdwlist} % use to allow single space itemize*

\usepackage{color}
\definecolor{mycolor}{rgb}{0.0,0.0,0.5}%dark blue
\usepackage{hyperref}
\hypersetup{colorlinks,breaklinks,
	linkcolor=mycolor,
	urlcolor=mycolor,
	anchorcolor=mycolor,
	citecolor=mycolor}

\newcommand{\dptdpt}{$\langle \delta p_{t1} \delta p_{t2}  \rangle$}
\newcommand{\R}{${\cal R}$}
\newcommand{\C}{${\cal C}$}
\newcommand{\D}{${\cal D}$}
\newcommand{\pt}{$p_t$}
\newcommand{\apt}{$\langle p_t\rangle$}
\newcommand{\pp}{$pp$}

\newcommand{\NR}{$\langle N\rangle {\cal R}$}
\newcommand{\NC}{$\langle N\rangle {\cal C}$}
\newcommand{\ND}{$\langle N\rangle {\cal D}$}
\newcommand{\Ndptdpt}{$\langle N\rangle(1+{\cal R}) \langle \delta p_{t1} \delta p_{t2}  \rangle$}

\usepackage{color}
\begin{document}

\title{Complementary Two-Particle Correlation Observables for Relativistic Nuclear Collisions}

\author{Mary Cody}
\email[]{mcody@ltu.edu}
\affiliation{Department of Natural Sciences, Lawrence Technological University, 21000 West Ten Mile Road, Southfield, MI  48075}

\author{Sean Gavin}
\email[]{sean.gavin@wayne.edu}
\affiliation{Department of Physics and Astronomy, Wayne State University, Detroit, MI, 48202}

\author{Brendan Koch}
\email[]{brendan.koch@duke.edu}
\affiliation{Department of Natural Sciences, Lawrence Technological University, 21000 West Ten Mile Road, Southfield, MI  48075}
\affiliation{Duke University Medical Center, Durham, NC, USA}

\author{Mark Kocherovsky}
\email[]{mkocherov@ltu.edu }
\affiliation{Department of Natural Sciences, Lawrence Technological University, 21000 West Ten Mile Road, Southfield, MI  48075}

\author{Zoulfekar Mazloum}
\email[]{zoulfekar.mazloum@wayne.edu}
\affiliation{Department of Physics and Astronomy, Wayne State University, Detroit, MI, 48202}

\author{George Moschelli}
\email[]{gmoschell@ltu.edu}
\affiliation{Department of Natural Sciences, Lawrence Technological University, 21000 West Ten Mile Road, Southfield, MI  48075}

\date{\today}

\begin{abstract}
Two-particle correlations are a widely used tool for studying relativistic nuclear collisions. Multiplicity fluctuations comparing charge and particle species have been studied as a possible signal for Quark-Gluon Plasma (QGP) and the QCD critical point. These fluctuation studies all make use of particle variances which can be shown to originate with a two-particle correlation function. Momentum correlations and covariances of momentum fluctuations, which arise from the same correlation function, have also been used to extract properties of the nuclear collision medium such as the shear viscosity to entropy density ratio, the shear relaxation time, and temperature fluctuations. Searches for critical fluctuations are also done with these correlation observables. We derive a mathematical relationship between several number and momentum density correlation observables and outline the different physics mechanisms often ascribed to each. This set of observables also contains a new multiplicity-momentum correlation. Our mathematical relation can be used as a validation tool for measurements, as a method for interpreting the relative contributions of different physics mechanisms on correlation observables, and as a test for theoretical and phenomenological models to simultaneously explain all observables. We compare an independent source model to simulated events from PYTHIA for all observables in the set. 
% no room to say stuff about D in particular being zero GCE equilbrium and being non-zero and positve in pythia.
\end{abstract}

\maketitle

%%%%%%%%%%%%%%%%%%%%%%%%%%%%%%%%%%%%%%%%%%%%%%%%%%%%%%%%%%%%%%%%%%%%%%%%%%%%%%%%%%%%%%%%%%%%%%%
%		INTRODUCTION
%%%%%%%%%%%%%%%%%%%%%%%%%%%%%%%%%%%%%%%%%%%%%%%%%%%%%%%%%%%%%%%%%%%%%%%%%%%%%%%%%%%%%%%%%%%%%%%
\section{\label{sec:intro} Introduction}
We present a set of two-particle number density and transverse momentum correlation observables that each separately test different aspects of relativistic heavy-ion collisions, but are mathematically connected through a parent correlation function. Several observables in the set have previously been measured individually, but are rarely measured simultaneously under the same collision system, energy, and acceptance conditions. One observable, multiplicity-momentum correlations, is new. The mathematical connection between the observables allows any one to be described as a combination of the others, signaling the relative contributions of the physical mechanisms of each. This connection also poses a challenge for models to address experimental measurements of all observables simultaneously. In this paper, we outline the construction and interpretation of the individual observables, demonstrate their mathematical connection, and compare a simple independent source model to simulated data. 

Two-particle correlation observables are widely used to study aspects of relativistic heavy-ion collisions. 
Multiplicity fluctuations, \R, have been linked to centrality or volume fluctuations  and studied as a possible signal for Quark-Gluon Plasma (QGP) \cite{Jeon:2000wg,Asakawa:2000wh,Koch:2001zn,Adams:2003st,Zhou:2018fxx, Heiselberg:2000ti,Adcox:2002mm,Zaranek:2001di, Mrowczynski:1997kz,Adare:2008ns,Mukherjee:2016hrj, Becattini:2005cc, Gavin:2000cu,Gavin:2001uk, Mrowczynski:2001mm,Pruneau:2002yf}. 
We outline these aspects and the experimental measurement 
in Sec.~\ref{sec:R}. The dependence of \R~on volume fluctuations is also informed by its representation in an independent source model which is discussed in Appendix \ref{sec:ISM}. 
%Volume fluctuations are determined by the fluctuation of sources of particles from event to event.  

Transverse momentum correlations, in the form of a covariance of two different particle's traverse momentum fluctuation away from the global average, \dptdpt, have also been examined as a signature of critical fluctuations and linked to event-by-event temperature fluctuations \cite{Appelshauser:1999ft, Anticic:2003fd,Adams:2003uw,Adams:2005ka,Adamova:2003pz,Adams:2005aw,Adams:2006sg,Adcox:2002pa, Adler:2003xq, Abelev:2014ckr, Heckel:2015swa,Adam:2017ucq,Acharya:2018ddg,Adam:2019rsf}. In past work, we argue that these correlations result from initial state correlations modified by radial flow \cite{Gavin:2011gr}. We also argue that these correlations can signal the level of thermalization reached by the collision medium \cite{Gavin:1990up,Gavin:2003cb,Gavin:2016nir}. 
We discuss this observable in detail in Sec.~\ref{sec:dptdpt}. 

We distinguish the net correlation of transverse momentum fluctuations, \dptdpt, from a different measure of two-particle transverse momentum correlations that we label \C. 
In \dptdpt, the fluctuation of an individual particle's transverse momentum away from the global average, $\delta p_t = p_{t}-\langle p_t\rangle$, is compared to that of every other particle. \C~directly compares two different particles' transverse momentum. 
\C~was first defined in Ref. \cite{Gavin:2006xd} and used there to estimate the shear viscosity to entropy density ratio and shear relaxation time  \cite{Agakishiev:2011fs,Acharya:2019oxz,Gonzalez:2020gqg,Gonzalez:2020bqm,Gonzalez:2018cty,Magdy:2021sip,Gavin:2016hmv,Moschelli:2018ntx,Moschelli:2019otk}. 
In this work, we focus on the mathematical and observable construction of these correlations and discuss how the effects of dynamical correlations and the influence of multiplicity fluctuations can be distinguished (see Secs.~\ref{sec:correlations} and~\ref{sec:C}). 
 
A major component of this work is the definition of a new two-particle correlation that measures the covariance of event-by-event multiplicity and total transverse momentum in excess of random fluctuations,  
\begin{equation}\label{eq:Dvar}
	{\cal D} =
	%\langle P_T N\rangle - \langle P_T\rangle\langle N\rangle - 
	%\langle p_t\rangle \left( \langle N^2\rangle -\langle N\rangle^2 \right)
	\frac{ Cov(P_T,N)-\langle p_t\rangle Var(N) }{\langle N\rangle^2}.
\end{equation}
The first mention of (\ref{eq:Dvar}) can be found in Ref.~\cite{Zin:2017awm}. 
In Sec.~\ref{sec:correlations}, we show that \D~emerges as a moment of the same parent correlation function that produces correlations \R, \C, and \dptdpt. 
More detailed discussion of the construction of \D~can be found in Sec.~\ref{sec:D}. 

In Sec.~\ref{sec:D} we discuss how \D~will vanish if the only source of multiplicity-momentum correlations is multiplicity fluctuations. Additionally we show that, in the Grand Canonical Ensemble, \D~is also zero in equilibrium. Non-zero values of \D~represent correlations born from the particle production mechanism that survive to freeze-out, and similarly to \dptdpt~\cite{Gavin:2016nir}, could be a sign of incomplete thermalization. 

Interestingly, in Sec.~\ref{sec:results}, we find that \D~is not zero in PYTHIA/Angantyr simulations of proton-proton (\pp) and nucleus-nucleus ($AA$) collisions. Also, we find \D~is comparable in magnitude to correlations of transverse momentum fluctuations, \dptdpt, which have been well measured at both RHIC and LHC.
Until this work, we have previously assumed \D~is zero and this is also assumed in Ref. \cite{Acharya:2019oxz} where ALICE measures two-particle transverse momentum correlations 
%${G}_2={\cal C}/\langle p_t\rangle^2$ 
differentially in relative pseudorapidity and relative azimuthal angle. 

The main result of this work is that multiplicity-momentum correlations, \D, net correlations of transverse momentum fluctuations, \dptdpt, multiplicity fluctuations, \R,  and transverse momentum correlations, \C, are mathematically related by the equation 
\begin{equation}\label{eq:sumrule}
(1+{\cal R})\langle \delta p_{t1} \delta p_{t2}\rangle 
- {\cal C} 
+ 2\langle p_t\rangle{\cal D} 
+ \langle p_t\rangle^2{\cal R} = 0.
\end{equation}
We derive this result in Sec.~\ref{sec:sumrule}. 
When each observable is measured individually, (\ref{eq:sumrule}) provides a previously unknown validation. Additionally, theoretical and phenomenological models that demonstrate good agreement with one observable can now use that comparison as a benchmark for simultaneously addressing the other observables. Importantly, each observable potentially represents a different physics effect; with (\ref{eq:sumrule}), one observable can be decomposed into the contributions from each different effect. 

In Sec.~\ref{sec:results}, we compare an Independent Source Model (ISM) to simulated data generated with PYTHIA 8.2 \cite{Sjostrand:2014zea} for several collision energies in \pp~collisions, as well as in $AA$ collisions using the Angantyr model \cite{Bierlich:2018xfw}. 
A deviation of experimental measurement from the ISM may signal novel physics. 
%such as incomplete thermalization or the existence of critical fluctuations.
The Angantyr model for heavy-ion collisions uses a superposition of nucleon-nucleon sub-collision model, similar to the wounded nucleon model, and allows for fluctuating positions of nucleons in the target and projectile nuclei. Additionally, multi-patron interactions and fluctuations exist in individual nucleon-nucleon sub-collisions. PYTHIA and Angantyr do not include any mechanism for collective expansion in \pp~or $AA$ collisions, therefore our calculations from simulated events provide a good comparison to ISM results.  Furthermore, the lack of collective effects makes the PYTHIA/Angantyr results an important baseline for experimental measurement. 
Using PYTHIA/Angantyr, we demonstrate the relationship (\ref{eq:sumrule}), and calculate the first estimate of \D, which we find to be non-zero and positive. 

%%%%%%%%%%%%%%%%%%%%%%%%%%%%%%%%%%%%%%%%%%%%%%%%%%%%%%%%%%%%%%%%%%%%%%%%%%%%%%%%%%%%%%%%%%%%%%%
%		CORRELATIONS AND FLUCTUATIONS
%%%%%%%%%%%%%%%%%%%%%%%%%%%%%%%%%%%%%%%%%%%%%%%%%%%%%%%%%%%%%%%%%%%%%%%%%%%%%%%%%%%%%%%%%%%%%%%
\section{Correlations and Fluctuations}\label{sec:correlations}
The construction of two-particle correlation observables begins by defining the two-particle momentum density 
\begin{equation}\label{eq:2pMomDis}
\rho_2(\mathbf{p}_1,\mathbf{p}_2) = 
\rho_1(\mathbf{p}_1)\rho_1(\mathbf{p}_2) +
r(\mathbf{p}_1, \mathbf{p}_2).
\end{equation}
Here $\mathbf{p}_{1,2}$ is the three-momentum of particle 1 or 2 in the pair. 
Single particle and pair momentum densities, $\rho_1(\mathbf{p})$ and $\rho_2(\mathbf{p}_1,\mathbf{p}_2)$, are the momentum densities of particles for an ensemble of events such that
\begin{eqnarray}
\rho_1(\mathbf{p}) &=& 
\frac{dN}{d^3\mathbf{p}}, \label{eq:rho1} \\
\rho_2(\mathbf{p}_1,\mathbf{p}_2) &=& 
\frac{dN}{d^3\mathbf{p}_1 d^3\mathbf{p}_2}, \label{eq:rho2}
\end{eqnarray}
and 
\begin{eqnarray}
\langle N\rangle &=& \int \rho_1(\mathbf{p})d^3\mathbf{p} \label{eq:avgN} \\
\langle N(N-1)\rangle &=& \iint \rho_2(\mathbf{p}_1,\mathbf{p}_2)d^3\mathbf{p}_1 d^3\mathbf{p}_2. \label{eq:avgPairs}
\end{eqnarray}
The angled brackets represent an average over the events in the ensemble. For any quantity $X$, the event average is defined as  $\langle X\rangle = N_{events}^{-1}\sum_{k=1}^{N_{events}} X_k$. 
Then, $\langle N\rangle$ is the average number of particles per event, and $\langle N(N-1)\rangle$ is the average number of particle pairs neglecting autocorrelations.

Equation (\ref{eq:2pMomDis}) highlights that particle pairs have two contributions. First, if pairs are formed from independent particles, i.e. no correlations, then the pair distribution is simply the multiplication of two single particle densities $\rho_1\rho_1$. Second, correlated pairs are represented by 
\begin{equation} \label{eq:CorrFn}
r(\mathbf{p}_1,\mathbf{p}_2) = \rho_2(\mathbf{p}_1,\mathbf{p}_2) - \rho_1(\mathbf{p}_1)\rho_1(\mathbf{p}_2).
\end{equation}
By construction, correlations vanish in the case of uncorrelated particle emission, when only statistical fluctuations are present. 

We focus on four moments of this distribution due to the complementary relationship provided by (\ref{eq:sumrule}),
\begin{eqnarray}
	{\cal R} &=& \frac{\iint r(\mathbf{p}_1,\mathbf{p}_2)d^3\mathbf{p}_1d^3\mathbf{p}_2}{\langle N\rangle^2}, \label{eq:Rint} \\[5pt]
	{\cal C} &=& \frac{\iint r(\mathbf{p}_1,\mathbf{p}_2)\, p_{t,1}p_{t,2}\, d^3\mathbf{p}_1d^3\mathbf{p}_2}{\langle N\rangle^2}, \label{eq:Cint} \\[5pt]
	\hspace{-9mm}\langle\delta p_{t1} \delta p_{t2}\rangle &=& 
	\frac{\iint r(\mathbf{p}_1,\mathbf{p}_2)\, \delta p_{t,1}\delta p_{t,2}\, 
		d^3\mathbf{p}_1 d^3\mathbf{p}_2}{\langle N(N-1)\rangle}, \label{eq:dptdptcorr}\\[5pt]
	&\text{and}&\nonumber\\[5pt]
	{\cal D} &=& \frac{\iint r(\mathbf{p}_1,\mathbf{p}_2)\, \delta p_{t,1}\, d^3\mathbf{p}_1d^3\mathbf{p}_2}{\langle N\rangle^2}. \label{eq:Dint} 
%	\\[5pt]
%	%
%	(1+&{\cal R}&)\langle \delta p_t \delta p_t\rangle 
%	- {\cal C} 
%	+ 2\langle p_t\rangle{\cal D} 
%	+ \langle p_t\rangle^2{\cal R} = 0. \label{eq:sumrule}
%	%
\end{eqnarray}
%
%Here, \R, (\ref{eq:Rint}), represents multiplicity fluctuations.  
%\C, (\ref{eq:Cint}), is the transverse momentum weighted version of \R~and represents transverse momentum correlations. 
%\dptdpt, (\ref{eq:dptdptcorr}), is the net correlation of transverse momentum fluctuations where 
Here 
\begin{equation}\label{eq:dpt}
	\delta p_{t,i} = p_{t,i}-\langle p_t\rangle
\end{equation}
is the fluctuation of the transverse momentum of particle $i$ %in event $k$ 
from the global average transverse momentum per particle, \apt, for a given centrality class.  
%Finally \D, (\ref{eq:Dint}), represents an excess multiplicity-momentum correlation. 

%%
%% flow arguement
%%
At this point, we make no assumption about the physical mechanisms that produce the correlations characterized by  (\ref{eq:CorrFn}), though many possibilities have been identified. 
Correlations of particle emission (momentum) angle with respect to an event plane is commonly called flow and is characterized by the  coefficients $v_n$ of a Fourier fit to the particle azimuthal distribution $dN/d\phi = \sum_{n} v_n \cos(n\phi -n\psi_n)$ \cite{Voloshin:1994mz, Borghini:2000sa, Voloshin:2008dg}. 
Since the event plane angle $\psi_n$ is calculated for each order and is identified with a geometrical shape -- $n=2$ is elliptical, $n=3$ is triangular, etc.~-- flow correlations are often called geometrical correlations. 
Much effort has gone into identifying so-called ``non-flow'' correlations that include HBT-like femtoscopic correlations \cite{Dinh:1999mn, Lisa:2005dd}, resonance decays and final state interactions \cite{Borghini:2000cm}, momentum conservation \cite{Danielewicz:1987in,Borghini:2007ku}, and jets. 
In other works, \cite{Gavin:2012if,Gavin:2011gr, Gavin:2008ev}, we have proposed that particles created in close spatial proximity develop a momentum correlation due to transverse expansion. We have argued that this mechanism accounts for much of the signal of two-particle correlations. Since this effect is only indirectly tied to an event reaction plane, many would label this effect as a non-flow effect. 

Instead of trying to diagnose the relative contributions from different correlation mechanisms in one observable, we propose the collection of observables 
(\ref{eq:Rint}), (\ref{eq:Cint}), (\ref{eq:dptdptcorr}), and (\ref{eq:Dint}) 
that all originate with (\ref{eq:2pMomDis}). Each observable has a different sensitivity to particle production, initial state correlations, and dynamical evolution; their mathematical connection, (\ref{eq:sumrule}), challenges experiments to measure each in a consistent way and constrains models to agree with all of the observables simultaneously. 
We comment that any theoretical model with a well-defined two-particle correlation function should produce correlation moments \eqref{eq:Rint}-\eqref{eq:Dint} that satisfy \eqref{eq:sumrule} since \eqref{eq:sumrule} is a mathematical identity. However, if a model is tuned to address experimental results for one observable, that tuning can now be tested by simultaneous comparison to all of \eqref{eq:Rint}-\eqref{eq:Dint}. This highlights the importance of simultaneous experimental measurements of \eqref{eq:Rint}-\eqref{eq:Dint} that satisfy \eqref{eq:sumrule}.

It is common in modern studies \cite{Acharya:2018ddg, Acharya:2019oxz, Agakishiev:2011fs} 
for experiments to measure correlation and fluctuation quantities differentially in relative azimuthal angle $\Delta\phi=\phi_1-\phi_2$ and relative pseudorapidity $\Delta \eta = \eta_1-\eta_2$. It is also common for experiments to measure pairs that are separated by a gap in pseudorapidity larger than $|\Delta\eta|\approx1$. 

Measuring observables as a function of $\Delta\phi$ allows for the diagnosis of contributions from anisotropic flow. 
Projections of differential measurements of   
%(\ref{eq:Rint}), (\ref{eq:Cint}), and (\ref{eq:dptdptcorr}) 
\R, \C, and \dptdpt~%
onto the $\Delta\phi$ axis all show a similar pattern of two peaks, one at $\Delta\phi=0$ and one at $\Delta\phi=\pi$ that is characteristic of both momentum conservation and anisotropic flow. These observables also show a broader peak at $\Delta\phi=\pi$ in comparison to the narrower peak at $\Delta\phi=0$. This observation is often attributed to the existence of triangular flow. 

Pseudorapidity gaps between pairs are used to eliminate ``short-range'', $|\Delta\eta|<1$, correlations such as resonance decays and jets. 
Separately, in differential measurements, HBT and track pileup effects are often removed by eliminating the $\Delta\eta=0$ bin. 
Projections onto the $\Delta\eta$ axis of the differential measurements of 
%(\ref{eq:Rint}), (\ref{eq:Cint}), (\ref{eq:dptdptcorr}), 
\R, \C, and \dptdpt~%
all show a ``long-range'', $|\Delta\eta|>1-2$, correlation in central collisions. This long-range ``near-side'' ($\Delta\phi=0$) correlation appears to extend beyond detector rapidity acceptances. Thus, the near-side peak is often described as a peak sitting on a long and flat pedestal, commonly called ``the ridge''. Experimental measurements often fit the ridge with a Fourier series like $\sum_{n}a_n\cos(n\Delta\phi)$ that is flat in $\Delta\eta$ and then relate the $a_n$ coefficients to the $v_n$ anisotropic flow harmonics  \cite{Magdy:2021sip,Acharya:2019oxz,Acharya:2018ddg,Alver:2010gr,Adam:2017ucq}. 
%\red{There must be other references that extract vn that I'm leaving out.} 
The peak sitting on the pedestal represents correlations in excess of the ridge (in excess of flow correlations) and still extends to long-range in $\Delta\eta$ (and possibly beyond the experimental acceptance) in central collisions. The broadness in $\Delta\eta$ of this excess decreases as collisions become more peripheral. Peripheral peaks have widths between $0.5-1$, which are consistent with jet and resonance decay correlations. The increasing $\Delta\eta$ width of the near side peak from peripheral to central collisions indicates that a correlation mechanism that is \textit{not} attributed to flow harmonics is at work. See, for example, Ref. \cite{Gavin:2016hmv}. 

If %the observables  
%(\ref{eq:Rint}), (\ref{eq:Cint}), (\ref{eq:dptdptcorr}), 
\R, \C, and \dptdpt~%
are not measured differentially in $(\Delta\eta,\Delta\phi)$, then all flow effects are eliminated. 
To understand this, imagine the quantity \R$(\Delta\phi)$ has been measured and is well described by a Fourier series with terms $a_n\cos(n\Delta\phi)$. To find the integrated quantity, one calculates \R$=\int_0^{2\pi} {\cal R}(\Delta\phi)d\Delta\phi$. When calculating the equivalent integral of the Fourier series, the integral of all terms $\cos(n\Delta\phi)$ over a symmetric interval vanish term by term, indicating ${\cal R}=0$ if correlations are only described by flow. Therefore, if the integrated quantity \R~is not zero, it is not fully explained by Fourier flow coefficients. Although these remaining correlations might be characterized as non-flow, they are still interesting and potentially provide useful information about the collision dynamics or initial state. 
In particular, we highlight the near-side correlations in excess of flow that are long-range in nature. 
Additionally, if jet effects dominate the integrated observables (\ref{eq:Rint})-(\ref{eq:Dint}), then analyzing them together may provide clues for identifying classes of events based on jet properties. Alternatively, the centrality and system energy dependence of these correlations can indicate the level of thermalization of events, which we leave to future work.

% ***
% *** The next two paragraphs are referenced in the results section as being at the end of Sec. IIA
% *** If you move these paragraphs, be sure to change that reference.
% ***
Correlations (\ref{eq:CorrFn}) also embody the event-by-event fluctuations in produced particles. Notice that integrating (\ref{eq:CorrFn}) over all momenta for both particles results in   
$\langle N(N-1)\rangle - \langle N\rangle^2 = Var(N) -\langle N\rangle$.
Here the variance of particles, $Var(N) = \langle N^2\rangle -\langle N\rangle^2$, characterizes the fluctuation in produced particles. If each event is independent of all other events, then this variance should follow Poisson statistics -- where the variance is equal to the mean -- resulting in a vanishing integral of (\ref{eq:CorrFn}). 

Non-Poissionian fluctuations indicate that a physical mechanism -- in initial state production, the dynamical expansion, or final state interactions -- generates fluctuations in a correlated way in all of the events of the ensemble; in this case $r(\mathbf{p}_1,\mathbf{p}_2)\neq 0$. Consequentially, since these fluctuations are tied to physical processes, they are not completely random and can be identified with correlation observables. Non-Poissionian behavior is seen in experiments as well as simulations, and will be discussed in the next sections.

%%%%%%%%%%%%%%%%%%%%%%%%%%%%%%%%%%%%%%%%%%%%%%%%%%%%%%%%%%%%%%%%%%%%%%%%%%%%%%%%%%%%%%%%%%%%%%%
%		R
%%%%%%%%%%%%%%%%%%%%%%%%%%%%%%%%%%%%%%%%%%%%%%%%%%%%%%%%%%%%%%%%%%%%%%%%%%%%%%%%%%%%%%%%%%%%%%%
\subsection{Multiplicity Fluctuations}\label{sec:R}
%
%Multiplicity fluctuations have been widely studied with the goal of signaling the onset of QGP.   
%Net charge fluctuations are used to distinguish QGP from hadron gas \cite{Jeon:2000wg,Asakawa:2000wh,Koch:2001zn,Adams:2003st}.   
%Such studies rely on the idea of ``volume fluctuations'' to connect event selections based on multiplicity to a geometric picture of the collision region \cite{Zhou:2018fxx}. 
%Other net charge fluctuation studies look for large divergences that could signal a QGP phase transition \cite{Heiselberg:2000ti,Adcox:2002mm,Zaranek:2001di}. 
%Inclusive multiplicity fluctuations have been linked to the isothermal compressibility of the system \cite{Mrowczynski:1997kz,Adare:2008ns,Mukherjee:2016hrj} assuming the mid-rapidity region can be described by the Grand Canonical Ensemble. 
%(A study intended to be used as a baseline of statistical fluctuations emerging from a hadron-resonance gas in the Canonical, Micro Canonical, and Grand Canonical Ensembles can be found here \cite{Becattini:2005cc}.) 
%Net baryon fluctuations are used to identify small regions of chiral condensates to classify events that signal QGP formation \cite{Gavin:2000cu,Gavin:2001uk}.
%%
%All of these references use observables constructed of moments of inclusive or identified particle multiplicities \cite{Mrowczynski:2001mm,Pruneau:2002yf}. 

In this section we outline aspects of the multiplicity fluctuation observable (\ref{eq:Rint}) that is measurable as 
\begin{equation}\label{eq:R}
	{\cal R} = 
	\frac{\langle N(N-1)\rangle - \langle N\rangle^2}{\langle N\rangle^2}
	= \frac{Var(N) - \langle N\rangle}{\langle N\rangle^2},
\end{equation}
\cite{Pruneau:2002yf}. 
%We discuss how \R~sets an overall scale for any two-particle correlation that can be derived from the correlation function (\ref{eq:CorrFn}). Due to this connection, we examine how the construction of \R~yields a characteristic $1/\langle N\rangle$ behavior that influences the interpretation of every two-particle correlation observable included in this paper.
%
From (\ref{eq:Rint}) we see that \R~is a direct integral of the correlation function (\ref{eq:CorrFn}) and therefore sets an overall scale for all two-particle correlations. There are three important features of \R~that influence the construction and interpretation of all observables in this work: 
First, is the choice of normalization of \R~by $\langle N\rangle^{-2}$. 
Second, is the non-Poissionian multiplicity distribution that makes \R~non-zero. This distribution has been measured as a Negative-Binomial Distribution \cite{Adare:2008ns}. 
Third, is the expected $\langle N\rangle^{-1}$ behavior with collision centrality.

In Ref. \cite{Pruneau:2002yf}, Pruneau et al. show that, for inclusive distributions, the observable \R~is robust against detection efficiency effects and acceptance limitations. 
To show this, we start by constructing (\ref{eq:R}) from the single particle distribution, $\rho_1$, and  pair distribution, $\rho_2$, using (\ref{eq:avgN}) and (\ref{eq:avgPairs}) and follow arguments from both Refs. \cite{Pruneau:2002yf} and \cite{Gavin:2011gr}. If (\ref{eq:avgN}) and (\ref{eq:avgPairs}) are given arbitrary normalizations $a$ and $b$ such that we have $\rho_2\rightarrow a\rho_2$ and $\rho_1\rho_1 \rightarrow b\rho_1\rho_1$, then (\ref{eq:R}) becomes 
\begin{equation}\label{eq:Rnorm}
	{\cal R}_{acc} = \frac{a}{b}{\cal R}+\frac{a-b}{b}.
\end{equation}
If $a\neq b$, then \R~will receive a scale and offset that could be detector and collision system and energy dependent. 
However, if $a$ and $b$ are equal, such as the case for detector tracking efficiency, then ${\cal R}={\cal R}_{measured}$. This motivates the choice to normalize \R~by $\langle N\rangle^{-2}$. For this reason, \C~was first constructed with the same normalization \cite{Gavin:2006xd}, and consequentially, in this work, we construct \D~in the same way. However, \dptdpt~was normalized to the number of particle pairs $\langle N(N-1)\rangle$, and this difference is the source of the $(1+{\cal R})$ factor in (\ref{eq:sumrule}). Since $(1+{\cal R})=\langle N(N-1)\rangle/\langle N\rangle^2$, it changes the normalization of \dptdpt~to match the other observables.

In the search for critical fluctuations, the PHENIX collaboration measured the scaled variance of the charged multiplicity 
\begin{equation}\label{eq:ScaledVar}
	\omega = \frac{\langle N^2\rangle-\langle N\rangle^2}{\langle N\rangle} = \frac{\sigma^2}{\mu}
\end{equation}
where $\langle N\rangle =\mu$ is the average charged particle multiplicity and $\sigma^2 = \langle N^2\rangle-\langle N\rangle^2$ is the variance \cite{Adare:2008ns}. 
The multiplicity distribution of heavy ion collisions follows a Negative Binomial Distribution (NBD) with mean $\mu$ and scaled variance $\omega = 1+\mu/k_{NBD}$, where $k_{NBD}$ is a parameter. The NBD parameter is related to (\ref{eq:R}) by
\begin{equation}\label{eq:RkNBD}
	{\cal R} = \frac{\sigma^2 -\mu}{\mu^2}=\frac{\omega - 1}{\mu}=\frac{1}{k_{NBD}}. 
\end{equation}
Importantly, subsets of a NBD, randomly sampled with constant probability, will have the same $k_{NBD}$. 

Using the properties of the NBD, we now show that it is acceptable to measure \R~using only a subset of the multiplicity; this guides our methods in Sec.~\ref{sec:results}. 
Let $\mu$ and $\omega$ be the mean multiplicity and scaled variance from an unlimited acceptance. Also let $\mu_{acc}$ and $\omega_{acc}$ be the mean and scaled variance from a fractional acceptance. 
%Defining the acceptance fraction as $f_{acc}=\mu_{acc}/\mu$, 
By definition, the scaled variance for fractional acceptance is then $\omega_{acc} = 1 + \mu_{acc}/k_{NBD}$. 
Using ${\cal R}=k_{NBD}^{-1}$ and the relation (\ref{eq:RkNBD}) for $\mu_{acc}$ and $\omega_{acc}$, we find ${\cal R} = (\omega_{acc} - 1)/\mu_{acc} = {\cal R}_{acc}$, since $k_{NBD}$ is identical for the full acceptance and fractional acceptance regions. 
This result is consistent with (\ref{eq:Rnorm}) for $a=b$, and makes \R~an ideal measure of the strength of correlations.

The variance of a NBD is proportional to the mean, but, importantly, always larger than the mean. In the case where $k_{NBD}\rightarrow\infty$, the NBD approaches the Poisson distribution where the variance equals the mean. 
Examining the rightmost definition of \R~in (\ref{eq:R}), notice that 
for the case of independent particle production, the distribution would be Poisson and then ${\cal R}=0$. 
In the case where the variance of particles is proportional to the mean, then both terms in the numerator have a scale of $\langle N\rangle$, and \R~will follow a $1/\langle N\rangle$ behavior. 
For this reason we expect that all observables \R, \C, \D, and $(1+{\cal R}) \langle \delta p_{t1} \delta p_{t2}  \rangle$ will all follow $1/\langle N\rangle$, a centrality and collision energy dependent quantity. 
Furthermore, in Appendix \ref{sec:ISM}, we show that in an independent source model, all of the observables trend like $\langle K\rangle^{-1}$ where $K$ is the number of sources in an event. This $1/\langle N\rangle$ or $1/\langle K\rangle$ behavior is a defining characteristic of these correlations and therefore we look for deviations from this trend to signal novel physics. 
To have a positive \R, the multiplicity variance must be larger than $\langle N\rangle$. To deviate from a $1/\langle N\rangle$ behavior, the multiplicity variance must also change faster or slower than $\langle N\rangle$ with increasing centrality. 

When investigating the centrality dependence of multiplicity fluctuations (\ref{eq:R}), biases are introduced 
if the same particles are used to measure correlations and to measure centrality. This will be discussed in detail in Sec.~\ref{sec:results}, however, it is informative to briefly discuss one aspect here. Imagine (\ref{eq:R}) was calculated from an ensemble of events where each event has exactly the same number of particles. Then, $\langle N^2\rangle = \langle N\rangle^2$ and 
\begin{equation}\label{eq:OneOverN}
	{\cal R}\rightarrow -\frac{1}{\langle N\rangle}.
\end{equation}
This shows a limiting behavior that is a response to multiplicity binning. 
%the absence of fluctuations that is present in all similarly constructed correlation observables. 
To avoid this effect, the multiplicity used to measure centrality must be different from the multiplicity used to calculate (\ref{eq:R}). 
This is acceptable because \R~is robust against acceptance effects, as discussed earlier in this section. 

%%%%%%%%%%%%%%%%%%%%%%%%%%%%%%%%%%%%%%%%%%%%%%%%%%%%%%%%%%%%%%%%%%%%%%%%%%%%%%%%%%%%%%%%%%%%%%%
%		C
%%%%%%%%%%%%%%%%%%%%%%%%%%%%%%%%%%%%%%%%%%%%%%%%%%%%%%%%%%%%%%%%%%%%%%%%%%%%%%%%%%%%%%%%%%%%%%%
\subsection{Transverse Momentum Correlations}\label{sec:C}
Two-particle transverse momentum correlations, (\ref{eq:Cint}), are measurable as
\begin{equation}\label{eq:C}
	{\cal C} = \frac{
		\left\langle \sum\limits^{N_k}_{i=1}\sum\limits^{N_k}_{j\neq i} p_{t,i} p_{t,j} \right\rangle
		- \langle P_T\rangle^2 }{\langle N\rangle^2},
\end{equation}
where
\begin{equation}\label{eq:avgTotPt}
	\langle P_T\rangle = \left\langle \sum\limits^{N_k}_{i=1} p_{t,i} \right\rangle 
	= \int \rho_1(\mathbf{p})\, p_t\, d^3\mathbf{p} 
\end{equation}
is the average total transverse momentum per event.  
 
% The momentum correlation observable (\ref{eq:C}) was first defined in Ref. \cite{Gavin:2006xd} as part of a method for extracting the shear viscosity to entropy density ratio, $\eta/s$, independently from flow harmonic measurements. STAR measured \C, for the first time, differentially in relative pseudorapidity and relative azimuhtal angle \C$(\Delta\eta,\Delta\phi)$ \cite{Agakishiev:2011fs}. This measurement constrained $\eta/s$  to a range of $0.06 < \eta/s < 0.21$ which is in agreement with hydrodynamic flow estimates and the predicted AdS/CFT lower limit of $\eta/s=1/4\pi$ \cite{Kovtun:2004de}. The measured range is due mostly to experimental systematic uncertainty which may be reducible by measuring the integrated form of the rapidity width of (\ref{eq:C}) like $\sigma^2_{\cal C} = \int {\cal C}(\Delta\eta)\Delta\eta^2d\Delta\eta$ without using any fit functions.  
% ALICE measures a slightly modified form of (\ref{eq:C}) defined as $G_2={\cal C}/\langle p_t\rangle^2$ \cite{Acharya:2019oxz,Gonzalez:2020gqg,Gonzalez:2020bqm,Gonzalez:2018cty}. The differential form of $G_2$ was recently used in \cite{Magdy:2021sip} to extract harmonic Fourier coefficients in $\Delta\phi$ from simulated data and compare them to harmonic flow coefficients $v_n$ measured with the cumulant method and a pseudorapidty gap of $|\eta|=0.7$. 
 
Momentum correlations (\ref{eq:C}) are sensitive to both number density fluctuations as well as transverse momentum fluctuations; both are necessary to address the diffusion of transverse momentum fluctuations due to shear viscosity. 
Reference \cite{Gavin:2006xd} predicts that the simultaneous diffusion and dampening of initial state momentum fluctuations due to shear viscous forces results in the broadening of correlations \C~in relative rapidity over the collision lifetime. Since central collisions have longer lifetimes than peripheral ones, a centrality dependent measurement of the relative rapidity width of \C~should show a monotonic increase. 
This behavior was first seen by STAR when they measured (\ref{eq:C}) differentially in relative pseudorapidity and relative azimuthal angle \C$(\Delta\eta,\Delta\phi)$ \cite{Agakishiev:2011fs}. 
 
%STAR found a differential correlation structure similar to the ridge, \R$(\Delta\eta,\Delta\phi)$, with a near-side peak at $\Delta\phi=0$ that is broad in $\Delta\eta$ and an away-side peak at $\Delta\phi=\pi$ that is flat in $\Delta\eta$. The double peaks in $\Delta\phi$ are commonly seen as an indication of hydrodynamic flow because they can be (mostly) characterized by a Fourier cosine series. However, on the near-side, correlations in excess of the Fourier fit exist and are peaked at $\Delta\eta = \Delta\phi=0$. In peripheral collisions these excess correlations have a narrow profile in $\Delta\eta$ that is consistent with resonance decay or jet correlations. As collisions become more central,  the rapidity width of excess correlations increases in agreement with \cite{Gavin:2006xd}.
 
%Unexpectedly, Ref. \cite{Agakishiev:2011fs} found that the rapidity broadening of the near side of \C$(\Delta\eta,\Delta\phi)$ was not Gaussian in nature. Instead, central collisions had two peaks in $\Delta\eta$ with a local minimum at $\Delta\eta=0$. In Refs. \cite{Gavin:2016hmv,Moschelli:2018ntx,Moschelli:2019otk} we argue that the non-Gaussian broadening is a signal of causal diffusion that depends on both the shear viscosity and the shear relaxation time. 
 
We can see how (\ref{eq:C}) incorporates number density fluctuations by writing it in terms of the correlation function (\ref{eq:CorrFn}) to find (\ref{eq:Cint}). 
Comparing (\ref{eq:Cint}) to (\ref{eq:Rint}), notice that all multiplicity fluctuations in (\ref{eq:C}) are the same as those in (\ref{eq:R}), except they are weighted by transverse momentum. 
This is important because every particle carries some momentum, and therefore correlations and diffusion of particles necessarily implies correlations and diffusion of momentum. 
%In larger multiplicity events, more momentum pairs are possible. Higher multiplicity events also have longer lifetimes which allow for correlations to develop due to dynamic processes like geometric flow, but longer lifetimes also allow more time for equilibration which destroys correlations. 
 
%We discuss transverse momentum correlations with number density fluctuations removed in Sec.~\ref{sec:dptdpt}, but 
\C~probes the transfer of transverse momentum correlations between two points in the QGP from small rapidity separation to larger separations.
Imagine an event with a fluctuating initial state. 
%In a hydrodynamic or kinetic theory picture, hot and cold spots are deposited throughout the collision volume each with a different local temperature and energy density. 
Movement toward equilibrium is driven, in part, by viscous forces transferring energy density or momentum density or particle number density from higher temperature spots to lower ones. Interestingly, shear viscosity transports momentum perpendicular to the direction of flow, therefore shear viscosity spreads transverse momentum fluctuations (and therefore correlations) in the longitudinal direction up to approximately $1-2$ units in relative rapidity. 
 
Momentum correlations emerge form the initial state because pairs of particles are emitted from the same source and are generally subject to local enforcement of conservation laws. Since particles originate at the same spatial location, they experience roughly the same dynamics and can develop new correlations with each other and with the global event plane due to transverse expansion \cite{Voloshin:2003ud, Gavin:2008ev, Gavin:2011gr,Gavin:2012if}. If correlations exist over rapidity ranges of $|\Delta\eta|>1-2$ units then causality requires that they develop at the early stages of the collisions \cite{Dumitru:2008wn}.

If momentum correlations originate because pairs of particles are emitted from the same source, then the number of correlated pairs is roughly proportional to the temperature of the source. The more pairs, the stronger the correlation. In equilibrium, the distinction between different sources is destroyed, reducing the strength of the correlation. In this manner, \C~is sensitive to the equilibration process and can be used to estimate levels of partial thermalization \cite{Gavin:2016nir}.
 
%%%%%%%%%%%%%%%%%%%%%%%%%%%%%%%%%%%%%%%%%%%%%%%%%%%%%%%%%%%%%%%%%%%%%%%%%%%%%%%%%%%%%%%%%%%%%%%
%		<dptdpt>
%%%%%%%%%%%%%%%%%%%%%%%%%%%%%%%%%%%%%%%%%%%%%%%%%%%%%%%%%%%%%%%%%%%%%%%%%%%%%%%%%%%%%%%%%%%%%%%
\subsection{Covariance of Transverse Momentum Fluctuations}\label{sec:dptdpt}
%
% This observable was first measured by STAR in Ref. \cite{Adams:2005ka}
%
Transverse momentum correlations in excess of multiplicity fluctuations, defined by (\ref{eq:dptdptcorr}), have been widely studied as a possible signal for the existence of QGP 
%and as an indication of temperature fluctuations of the nuclear collision medium 
\cite{Appelshauser:1999ft, Anticic:2003fd,Adams:2003uw,Adams:2005ka,Adamova:2003pz,Adams:2005aw,Adams:2006sg,Adcox:2002pa, Adler:2003xq, Abelev:2014ckr, Heckel:2015swa,Adam:2017ucq,Acharya:2018ddg,Adam:2019rsf}. 
QCD critical point searches look for non-monotonic behaviors since fluctuations are expected to diverge if the system passes through a phase transition \cite{Heiselberg:2000ti,Rybczynski:2003jk}. 
%\red{Are those appropriate citations?} 
Similarly, the event-by-event variation in \pt~can be used as a measure of event temperature fluctuations \cite{Adams:2005ka,Tannenbaum:2001gs}. 

We focus on momentum correlations defined by (\ref{eq:dptdptcorr}), which are experimentally measurable with
\begin{equation}\label{eq:dptdpt}
\langle \delta p_{t1} \delta p_{t2}\rangle =
\frac{
	\left\langle \sum\limits^{N_k}_{i=1}\sum\limits^{N_k}_{j=1,j\neq i}\delta p_{t,i}\delta p_{t,j}\right\rangle
}
{\langle N(N-1)\rangle },
\end{equation}
where $\delta p_{t,i}$, defined by (\ref{eq:dpt}). Since (\ref{eq:dpt}) is a fluctuation, (\ref{eq:dptdpt}) is a covariance of fluctuations. In this work, we distinguish correlations of transverse momentum fluctuations (\ref{eq:dptdpt}) from transverse momentum correlations in Sec.~\ref{sec:C} %(\ref{eq:C}) 
to avoid confusion. We will discuss the relationship between these two types of momentum correlations in Sec.~\ref{sec:sumrule}. 

When two particles both have larger or smaller \pt~than the average, that pair contributes positively to \dptdpt. When one particle of a pair has positive $\delta p_t$ and the other has negative $\delta p_t$, then that pair contributes negatively to \dptdpt. In the case of purely independent particle emission, \dptdpt$=0$. 

\begin{figure}
	%\centering
	\includegraphics[width=\linewidth]{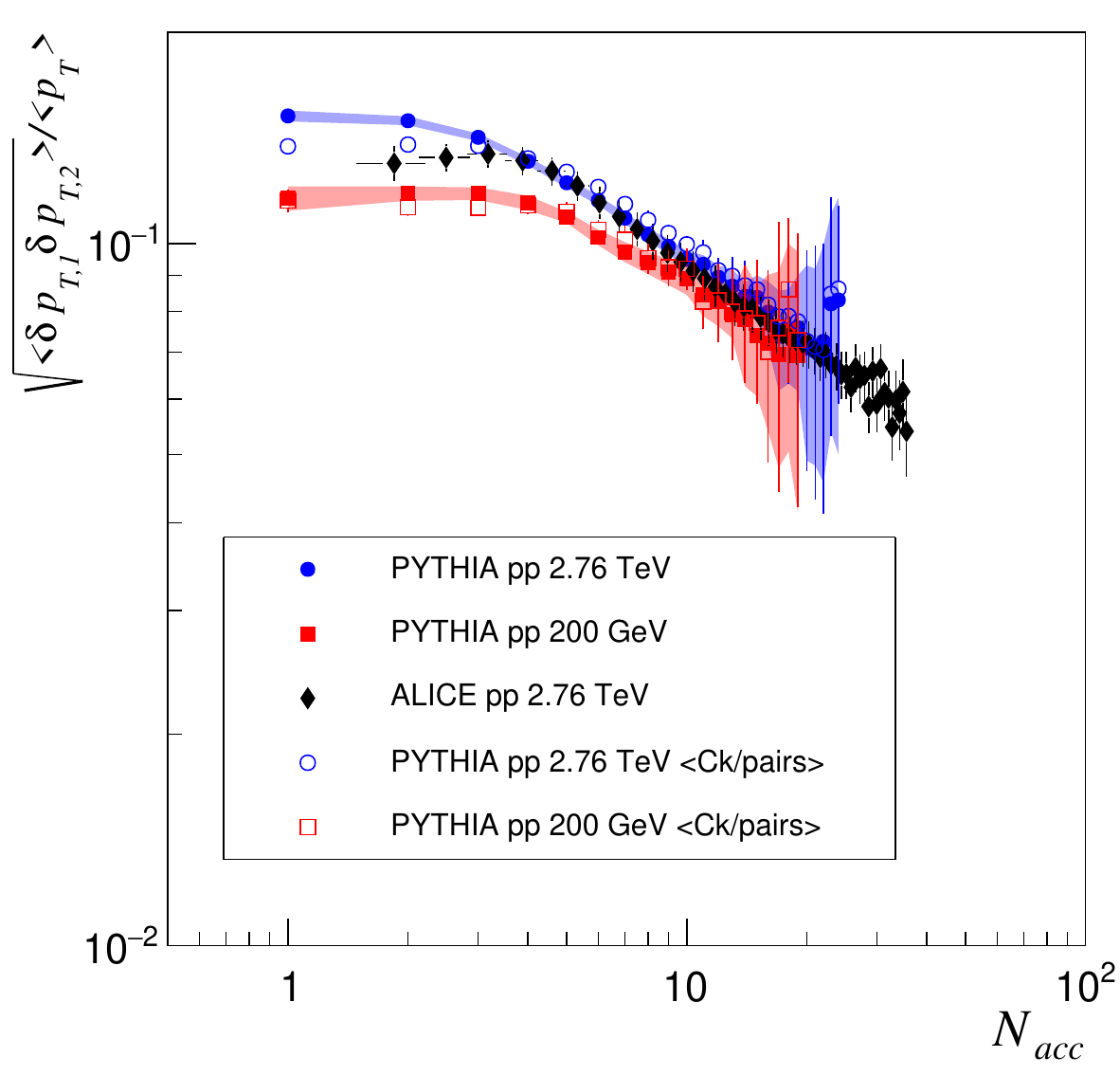}% sqrtCm_o_pt_pp % dptdpt_o_ptSq_pp
	\caption{Comparison of Eq.~(\ref{eq:sqrtCmOpt}) calculations from PYTHIA \pp~events (circles and squares) with measurement from ALICE (solid diamonds) \cite{Abelev:2014ckr, Heckel:2015swa}. Solid circles and squares represent (\ref{eq:dptdpt}), while open circles and squares represent (\ref{eq:dptdptSTAR}). }
	\label{fig:Cm_ptSq_pp}
\end{figure}

The definition (\ref{eq:dptdpt}) differs slightly from definitions found in experimental measurements. Experiments measure  
\begin{equation}\label{eq:dptdptSTAR}
\langle \delta p_{t1} \delta p_{t2}\rangle =
\frac{1}{N_{event}}\sum\limits_{k=1}^{N_{event}}\frac{C_k}{N_k(N_k-1)}
\end{equation}
with
\begin{equation}\label{eq:Cm}
C_k = \sum\limits_{i=1}^{N_{k}}\sum\limits_{j=1,j\neq i}^{N_{k}} (p_{t,i}-M_{p_t}\rangle)(p_{t,i}-M_{p_t})
\end{equation}
and
\begin{equation}\label{eq:avgpt_ebe}
M_{p_t} = \frac{1}{N_{event}}\sum\limits_{k=1}^{N_{event}}\langle p_t\rangle_k
\end{equation}
where $\langle p_t\rangle_k$ is the average transverse momentum in event $k$,
\begin{equation}\label{eq:avgpt_k}
\langle p_t\rangle_k = \frac{1}{N_{k}}\sum\limits_{i=1}^{N_{k}}p_{t,i}.
\end{equation}
There are two differences. First, the average transverse momentum (\ref{eq:avgpt_ebe}) is calculated event-by-event such that the average transverse momentum per particle of each event is found first then averaged over all events in the same centrality class. In (\ref{eq:dptdpt}) we define the average transverse momentum per particle as
\begin{equation}\label{eq:avgpt}
\langle p_t\rangle 
= \left\langle P_T\right\rangle/\left\langle N\right\rangle.
\end{equation}
where $\langle P_T\rangle$ is (\ref{eq:avgTotPt}), 
which is more representative of our theoretical description of the momentum density (\ref{eq:rho1}). 
The second difference between (\ref{eq:dptdpt}) and (\ref{eq:dptdptSTAR}) is in the normalization. The denominator of (\ref{eq:dptdpt}) is calculated independently, where the ratio $C_k / N_k(N_k-1)$ is calculated event-by-event in (\ref{eq:dptdptSTAR}). We make this choice in (\ref{eq:dptdpt}) to maintain as much consistency as possible between (\ref{eq:Rint}), (\ref{eq:Cint}), (\ref{eq:dptdptcorr}), and (\ref{eq:Dint}). 
In Fig.~\ref{fig:Cm_ptSq_pp} we plot both (\ref{eq:dptdpt}) and (\ref{eq:dptdptSTAR}) calculated with the same PYTHIA events. Excellent agreement is observed. 

%
%%%%%%%%%%%%%%%%%%%%%%%%%%%%%%%%%%%%%% FIG: <sqrtCm_pt> %%%%%%%%%%%%%%%%%%%%%%%%%%%%%%
\begin{figure}
	%\centering
	\includegraphics[width=\linewidth]{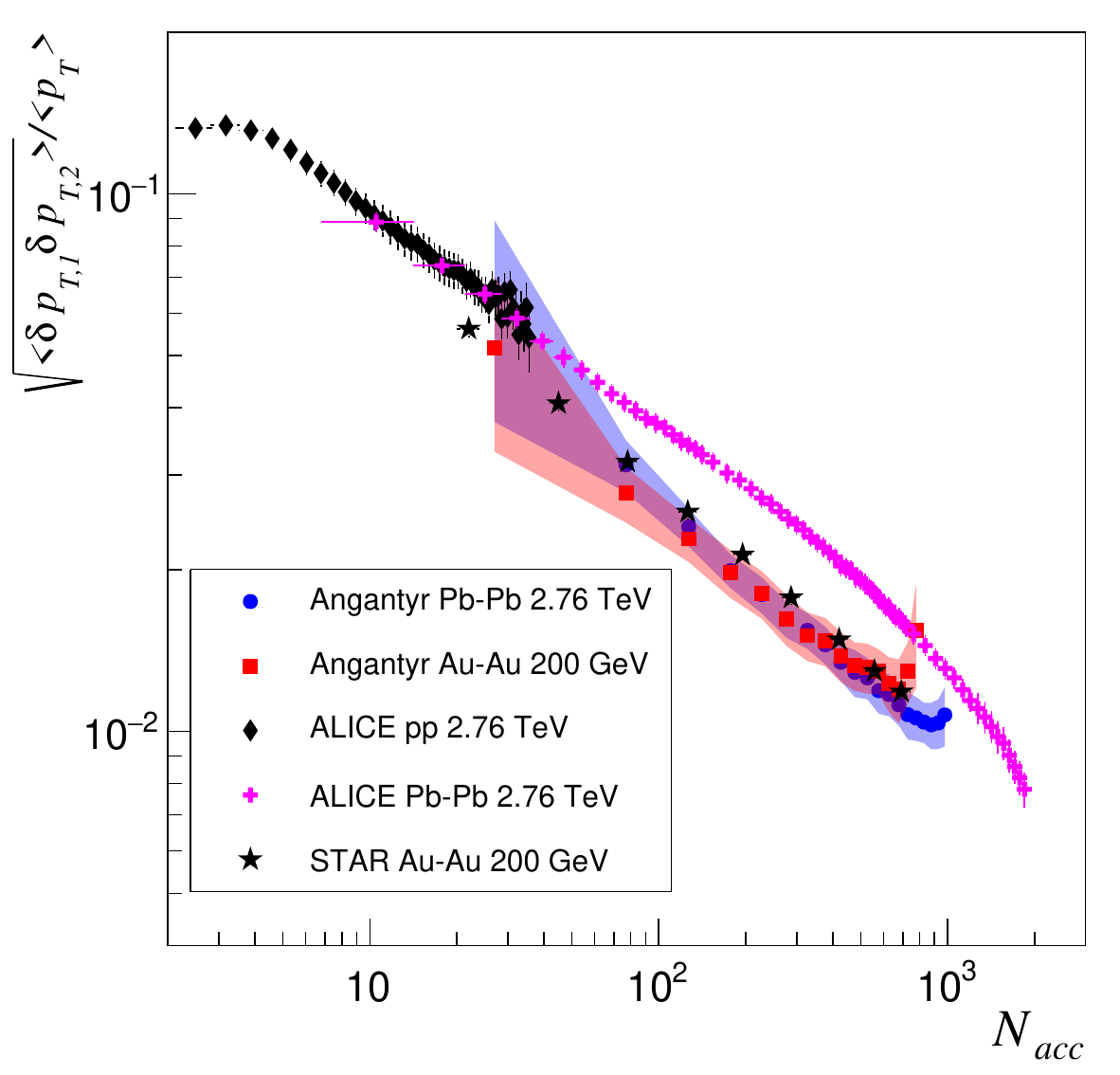}% sqrtCm_o_pt_AA % dptdpt_o_ptSq_AA
	\caption{Comparison of Eq.~(\ref{eq:dptdpt}) calculations from PYTHIA $AA$ events with measurement from ALICE \pp~and Pb-Pb collisions \cite{Abelev:2014ckr, Heckel:2015swa}, and STAR Au-Au collisions \cite{Adam:2019rsf}. Centrality is determined by multiplicity.}
	\label{fig:Cm_ptSq_AA}
\end{figure}
\begin{figure}
	%\centering
	\includegraphics[width=\linewidth]{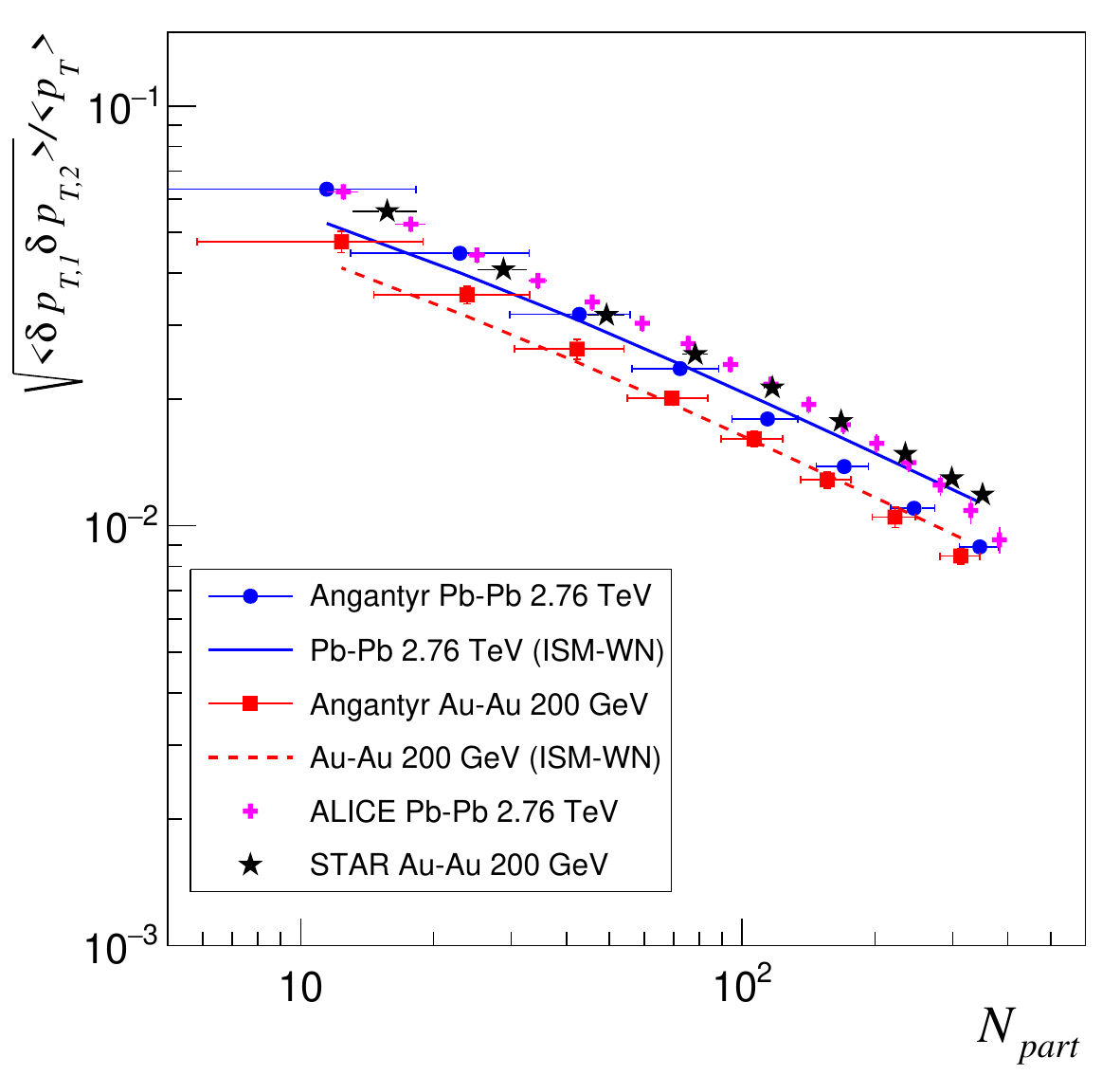}% sqrtCm_o_pt_Np % dptdpt_o_ptSq_Np
	\caption{Comparison of Eq.~(\ref{eq:dptdpt}) calculations from PYTHIA $AA$ events with measurement from ALICE \pp~and Pb-Pb collisions \cite{Abelev:2014ckr, Heckel:2015swa}, and STAR Au-Au collisions \cite{Adam:2019rsf}. Centrality is determined by the number of participating nucleons. Solid lines represent the independent source model for wounded nucleons, Eq.~(\ref{eq:dptdpt_WN}).}
	\label{fig:Cm_ptSq_Np}
\end{figure}

Experiments report positive values of \dptdpt~in \pp~and $AA$ collisions at various energies. \dptdpt~decreases with centrality, but not quite following $1/\langle N\rangle$ \cite{Adams:2005ka,Abelev:2014ckr,Adam:2019rsf}. If \dptdpt~falls with $1/\langle N\rangle$, then the quantity $(dN/d\eta)$\dptdpt~should be approximately flat. 
However, experimental measurements of $(dN/d\eta)$\dptdpt~rise from peripheral to mid-peripheral collisions and plateau toward more central collisions. This rise could signal the onset of critical fluctuations \cite{Adams:2005ka,Abelev:2014ckr} or the effects of incomplete thermalization \cite{Gavin:2016nir}.  

It is common for experimental measurements to report \dptdpt~as a relative dynamical correlation 
\begin{equation}\label{eq:sqrtCmOpt}
\sqrt{\langle \delta p_{t1} \delta p_{t2}\rangle}/\langle p_t\rangle,
%&\text{or}&\nonumber\\
%& \langle \delta p_{t1} \delta p_{t2}\rangle/\langle p_t\rangle^2&\label{eq:CmOptSq}
\end{equation}
which is dimensionless. It also rescales the growth of (\ref{eq:dptdpt}) to be dependent on the cumulative effect of correlations rather than the size of \apt. 
This scaling nearly removes the collision energy dependence of the measurements \cite{Adams:2005ka,Abelev:2014ckr,Adam:2019rsf}. 
Using PYTHIA/Angantyr simulated events, we calculate (\ref{eq:sqrtCmOpt}) using (\ref{eq:dptdpt}) and (\ref{eq:avgpt}), and compare to experimental data in Figs.~\ref{fig:Cm_ptSq_pp},~\ref{fig:Cm_ptSq_AA}, and~\ref{fig:Cm_ptSq_Np}. Details are discussed in Sec.~\ref{sec:results}. 

As with \R~and \C, experiments measure \dptdpt~differentially in relative rapidity and relative azimuthal angle $(\Delta\eta,\Delta\phi)$. The ALICE collaboration measures the quantity $P_2(\Delta\eta,\Delta\phi)=$~\dptdpt$(\Delta\eta,\Delta\phi)/\langle p_t\rangle^2$ which shows the characteristic ridge-like shape for charge independent correlations \cite{Acharya:2018ddg}. 
The near-side ridge at $\Delta\phi=0$ is not completely explainable with Fourier decomposition, and excess correlations appear to be long-range in $\Delta\eta$. 
Short range effects like resonance decays and jets can produce positive correlations in \dptdpt, but they cannot fully explain the excess long range correlations seen in $P_2(\Delta\eta,\Delta\phi)$. 

Several explanations for these correlations have been proposed. 
They include the quark coalescence models \cite{Xu:2020pxj}, string percolation models where clustered strings produce colored sources \cite{Ferreiro:2003dw}, fluctuations in event size and entropy \cite{Bozek:2017jog}, and a boosted source model where correlations originating in initial state hot-spots are enhanced by radial flow \cite{Gavin:2011gr}. 
We propose that any explanation of these correlations, should simultaneously address other two-particle correlations that originate from (\ref{eq:CorrFn}) such as (\ref{eq:Rint}), (\ref{eq:Cint}), and (\ref{eq:Dint}).  

%%%%%%%%%%%%%%%%%%%%%%%%%%%%%%%%%%%%%%%%%%%%%%%%%%%%%%%%%%%%%%%%%%%%%%%%%%%%%%%%%%%%%%%%%%%%%%%
%		D
%%%%%%%%%%%%%%%%%%%%%%%%%%%%%%%%%%%%%%%%%%%%%%%%%%%%%%%%%%%%%%%%%%%%%%%%%%%%%%%%%%%%%%%%%%%%%%%
\subsection{Covariance of Multiplicity and Transverse Momentum}\label{sec:D}
A new observable \D, defined by (\ref{eq:Dint}), tests the correlation of transverse momentum with particle production event-by-event. In Sec.~\ref{sec:results} we show that in PYTHIA/Angantyr simulations \D~is generally positive and comparable in magnitude to \dptdpt.

In (\ref{eq:Dint}), $\delta p_{t}$ is defined by (\ref{eq:dpt}),  
and \apt~is the average transverse momentum per particle for a given centrality class of events (\ref{eq:avgpt}). Experimentally, (\ref{eq:Dint}) can be measured with the final state particle pair sum
\begin{equation}\label{eq:D}
{\cal D} =
\frac{\left\langle \sum\limits^{N_k}_{i=1}\sum\limits^{N_k}_{j=1,j\neq i} \delta p_{t,i} \right\rangle}{\langle N\rangle^2}
= \frac{\left\langle (N_k-1) \sum\limits^{N_k}_{i=1} \delta p_{t,i} \right\rangle}{\langle N\rangle^2}. 
\end{equation}

To understand  this observable, we expand $\delta p_{t,i}$ in the middle term of (\ref{eq:D}) with (\ref{eq:dpt}) and substitute  
\begin{eqnarray}
\left\langle\sum\limits^{N_k}_{i=1}\sum\limits^{N_k}_{j=1,j\neq i} p_{t,i}\right\rangle &=& 
\langle P_T N\rangle - \langle P_T\rangle \label{eq:pt_pairs}\\
\text{and} \nonumber \\
\left\langle\sum\limits^{N_k}_{i=1}\sum\limits^{N_k}_{j=1,j\neq i} \langle p_{t}\rangle\right\rangle &=& \langle p_{t}\rangle\langle N(N-1)\rangle. \label{eq:aptNpairs}
\end{eqnarray}
Adding and subtracting $\langle P_T\rangle \langle N\rangle$ and making use of the fact that (\ref{eq:avgpt}) can be written as $\langle P_T\rangle = \langle p_t\rangle \langle N\rangle$, we find Eq.~(\ref{eq:Dvar}), where $Cov(P_T,N) = \langle P_T N\rangle - \langle P_T\rangle\langle N\rangle$ is the covariance of total transverse momentum $P_T$ and multiplicity $N$ per event.  The event multiplicity variance is $Var(N) = \langle N^2\rangle -\langle N\rangle^2$. 

Since every particle carries some transverse momentum, adding any particle to an event will increase the total transverse momentum in that event. Therefore, a natural correlation between total \pt~and multiplicity exists that is dominated purely by multiplicity fluctuations. 
Notice this contribution is subtracted by the rightmost term of (\ref{eq:Dvar}). This indicates that \D~should be zero if multiplicity fluctuations are the only source of multiplicity-momentum correlations.

In the Grand Canonical Ensemble, we can follow Ref. \cite{Zin:2017awm} to show that \D~should vanish in equilibrium. 
%\red{In this scenario, we must assume that the measured midrapidity region is in thermal contact with the remaining collision volume, which acts as a particle and heat reservoir.} 
In equilibrium, the Grand Partition Function with chemical potential $\mu$, volume $V$, and temperature $T$, is ${\cal Z}(\mu, V, T)=\sum_i \exp(\alpha N_i - \beta E_i)$. Here the Gibbs factor - with number of particles $N_i$ and energy $E_i$ of state $i$  - is summed over all states. We define $\alpha = \mu/T$ and $\beta = 1/T$ and take the Boltzmann constant to be in natural units $k_B=1$. The average number of particles and average energy are found in the usual way
\begin{eqnarray}
\langle N\rangle &=& \sum_i N_i\frac{e^{\alpha N_i - \beta E_i}}{\cal Z} =
\frac{1}{\cal Z}\frac{\partial {\cal Z}}{\partial \alpha}, \label{eq:GrandN} \\
\langle E\rangle &=& \sum_i E_i\frac{e^{\alpha N_i - \beta E_i}}{\cal Z} =
-\frac{1}{\cal Z}\frac{\partial{\cal Z}}{\partial \beta}. \label{eq:GrandE} 
\end{eqnarray}
Second derivatives in $\alpha$ yield
\begin{eqnarray}
\frac{\partial \langle N\rangle}{\partial\alpha} &=& 
\sum_i N_i \left( \frac{N_i e^{\alpha N_i - \beta E_i}}{\cal Z} - \frac{e^{\alpha N_i - \beta E_i}}{{\cal Z}^2}\frac{\partial {\cal Z}}{\partial \alpha} \right) \nonumber \\
&=& \langle N^2\rangle -\langle N\rangle^2, \label{eq:GandNsq}\\
~\nonumber \\
\frac{\partial \langle E\rangle}{\partial\alpha} &=& 
\sum_i E_i \left( \frac{N_i e^{\alpha N_i - \beta E_i}}{\cal Z} - \frac{e^{\alpha N_i - \beta E_i}}{{\cal Z}^2}\frac{\partial {\cal Z}}{\partial \alpha} \right) \nonumber \\
&=& \langle NE\rangle -\langle N\rangle\langle E\rangle = 
\frac{\partial \langle E\rangle}{\partial \langle N\rangle}\frac{\partial \langle N\rangle}{\partial \alpha}. \label{eq:GandNE}
\end{eqnarray}
Defining ${\cal D}_{E}=\left(Cov(E,N)-\varepsilon Var(N)\right)/\langle N\rangle^2$ where $\varepsilon =\langle E\rangle/\langle N\rangle$, we find that ${\cal D}_E$ vanishes when the energy per particle  satisfies $\varepsilon=\partial \langle E\rangle/\partial \langle N\rangle$.

To relate energy and transverse momentum fluctuations, we take the transverse mass to be $m_t = \sqrt{m^2 +p_t^2}\approx p_t$ for particles with large momentum, $p_t \gg m$. Near mid-rapidity $y\approx 0$, the energy $E_i=m_{t,i}\cosh y_i\approx m_{t,i}\approx p_{t,i}$ averaged over states is then approximately the average total transverse momentum $\langle E\rangle \approx \langle P_T\rangle$. Following that, we substitute $\partial\langle E\rangle/\partial\langle N\rangle = \partial\langle P_T\rangle/\partial\langle N\rangle$ in the last term of (\ref{eq:GandNE}). In the case where \apt~is constant over a wide range of multiplicities, the definition $\langle P_T\rangle = \langle p_t\rangle \langle N\rangle$, Eq.~(\ref{eq:avgpt}), yields 
$\partial\langle P_T\rangle/\partial\langle N\rangle \approx \langle p_t\rangle$. Using this result in (\ref{eq:GandNE}) with (\ref{eq:GandNsq}), we find 
\begin{equation}\label{eq:Dzero}
\langle NP_T\rangle -\langle N\rangle\langle P_T\rangle = \langle p_t\rangle(\langle N^2\rangle - \langle N\rangle^2).
\end{equation}
Finally, substituting (\ref{eq:Dzero}) in (\ref{eq:Dvar}), we find that \D$\,=0$.

%
%%%%%%%%%%%%%%%%%%%%%%%%%%%%%%%%%%%%%% FIG: <<pt>> %%%%%%%%%%%%%%%%%%%%%%%%%%%%%%
\begin{figure}
	%\centering
	\includegraphics[width=\linewidth]{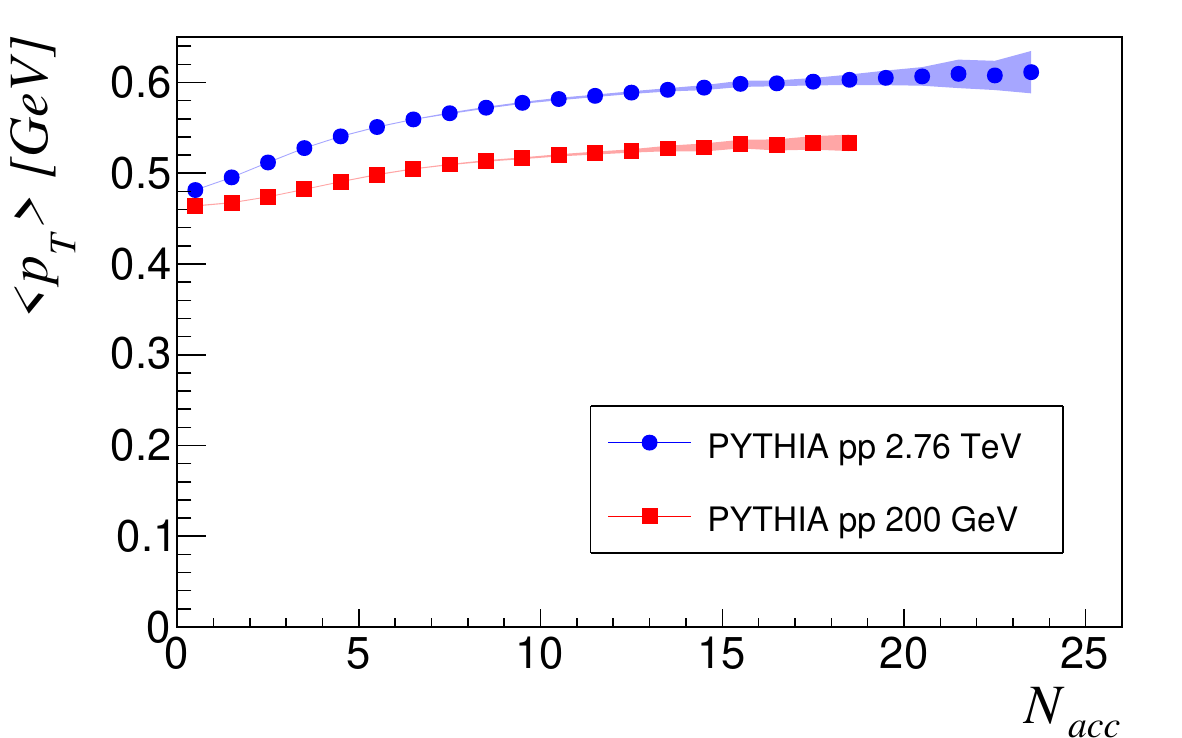}
	\caption{Average transverse momentum per particle for all charged particles as a function of reference multiplicity for \pp~collisions at select energies. Error bars on the PYTHIA results represent statistical uncertainty.}
	\label{fig:avgpt_pp}
	%\centering
	\includegraphics[width=\linewidth]{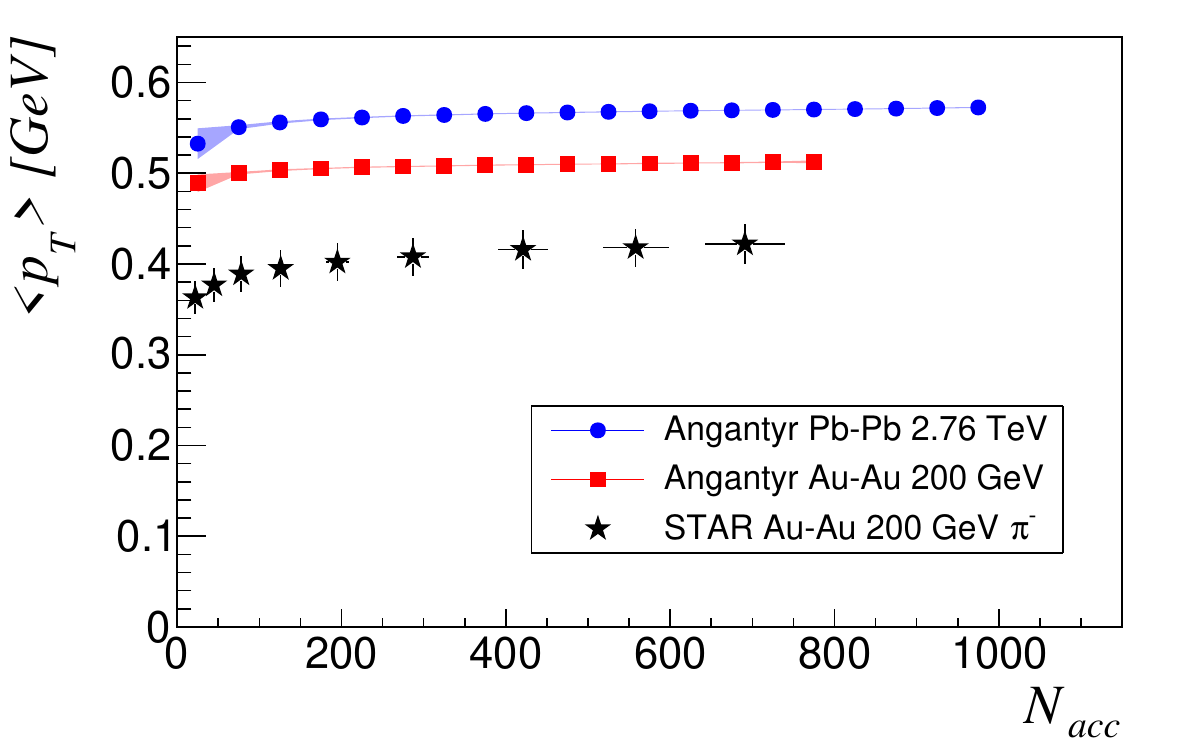}
	\caption{Average transverse momentum per particle for all charged particles as a function of reference multiplicity for select $AA$ collision systems. Error bars on the PYTHIA results represent statistical uncertainty. STAR data is from \cite{Abelev:2008ab}.}
	\label{fig:avgpt_AA}
\end{figure}

Several factors may generate a non-zero \D. 
Hadronization may violate the assumption that $p_t \gg m$ for all particles. For example, in $\sqrt{s}=200~GeV$ collision systems, the average transverse momentum is approximately \apt$\,\approx 0.5~GeV$, which is arguably large compared to the pion mass, but not the kaon or proton masses. Heavier particles may skew the momentum-multiplicity covariance $Cov(P_T,N)$. 
Correlations could be affected by canonical effects in the same way proton cumulants are influenced by baryon conservation (an interesting topic we leave for future work).  
Particle rapidities greater than $|y|>0.5$ have increasingly larger deviations from our $y=0$ assumption. 
If higher momentum particles, say with $p_t>2~GeV$, come at the cost of producing fewer particles near the average momentum, then the covariance $Cov(P_T,N)$ would become negative. If high momentum particles come in conjunction with excess particles near the average, then the covariance $Cov(P_T,N)$ will be positive. 
Lastly, the assumption $\partial\langle P_T\rangle/\partial\langle N\rangle \approx \langle p_t\rangle$ does not hold if transverse momentum per particle increases with increasing event multiplicity.

Examining \apt~vs. multiplicity, Figs. \ref{fig:avgpt_pp} and \ref{fig:avgpt_AA}, we notice that the average transverse momentum per particle increases with event multiplicity. This is seen by experiment across collision systems and energies. See, for example, Ref. \cite{Abelev:2013bla}. This is a positive transverse momentum and multiplicity covariance, if only a slight one. 
Possible sources of this covariance include jet particle production or an increased radial flow velocity in central collisions in comparison to peripheral collisions. 
This increase in \apt~has been seen in PYTHIA and is considered a consequence of the multiple interaction model \cite{Sjostrand:1987su} and color reconnection \cite{Sjostrand:2013cya}.
In all of these cases, non-zero \D~indicates a correlation related to particle production and dynamics that is distinct from \R, \C, and \dptdpt. We will show in Sec.~\ref{sec:sumrule}, how correlations \D~contribute to the other observables \C~and \dptdpt. 

%%%%%%%%%%%%%%%%%%%%%%%%%%%%%%%%%%%%%%%%%%%%%%%%%%%%%%%%%%%%%%%%%%%%%%%%%%%%%%%%%%%%%%%%%%%%%%%
%		SUM RULE
%%%%%%%%%%%%%%%%%%%%%%%%%%%%%%%%%%%%%%%%%%%%%%%%%%%%%%%%%%%%%%%%%%%%%%%%%%%%%%%%%%%%%%%%%%%%%%%
\subsection{Complimentary Fluctuation and Correlation Observables}\label{sec:sumrule}
The observables \R, (\ref{eq:R}), \C, (\ref{eq:C}), \dptdpt, (\ref{eq:dptdpt}), and \D, (\ref{eq:D}), are mathematically related based on their common origin (\ref{eq:CorrFn}) and the definition of a transverse momentum fluctuation $\delta p_t$, (\ref{eq:dpt}). We find the relation (\ref{eq:sumrule}).  

Starting from the definition (\ref{eq:dptdpt}) we expand the argument $\delta p_{t,i}\delta p_{t,j}$, to find 
\begin{eqnarray}
&&\langle \delta p_{t1} \delta p_{t2}\rangle = \nonumber \\
&&\frac{
	\left\langle 
	\sum\limits^{N_k}_{i=1}\sum\limits^{N_k}_{j\neq i}
	\left(
	p_{t,i}p_{t,j} - p_{t,i}\langle p_t\rangle - p_{t,j}\langle p_t\rangle +\langle p_t\rangle^2
	\right)
	\right\rangle
}
{\langle N(N-1)\rangle }. \label{eq:sumrule_start}
\end{eqnarray}
Applying the sums over pairs, we find 
\begin{subequations}\label{eq:sumrule_terms}
\begin{align}
\hspace{-5mm}\langle N(N-1)\rangle\langle \delta p_{t1} \delta p_{t2}\rangle &= \nonumber \\
&=
\langle N\rangle^2{\cal C} +\langle P_T\rangle^2
\label{eq:sumrule_C}\\
&- 2 (\langle P_T N\rangle -\langle P_T\rangle)\langle p_t\rangle \label{eq:sumrule_crossterms}\\
&+ \langle p_t\rangle^2\langle N(N-1)\rangle. \label{eq:sumrule_multfluc}
\end{align}
\end{subequations}
Using (\ref{eq:C}), the first term of (\ref{eq:sumrule_start}), becomes (\ref{eq:sumrule_C}). 
The middle two terms in (\ref{eq:sumrule_start}) both become $(\langle P_T N\rangle -\langle P_T\rangle)\langle p_t\rangle$ after using (\ref{eq:pt_pairs}), resulting in (\ref{eq:sumrule_crossterms}). 
The last term of (\ref{eq:sumrule_start}) yields (\ref{eq:sumrule_multfluc}), similarly to (\ref{eq:aptNpairs}). 
After adding and subtracting $2\langle p_t\rangle \langle P_T\rangle 
\langle N\rangle + 2\langle p_t\rangle^2\langle N^2\rangle$ to (\ref{eq:sumrule_terms}), we make use of definitions (\ref{eq:Dvar}) and (\ref{eq:R}) to construct 
\begin{equation}\label{eq:dptdpt_sumrule}
\langle \delta p_{t1} \delta p_{t2}\rangle 
=\frac{ {\cal C} - 2\langle p_t\rangle{\cal D} - \langle p_t\rangle^2{\cal R} }
{(1+{\cal R})},
\end{equation}
where $(1+{\cal R})=\langle N(N-1)\rangle/\langle N\rangle^2$. 

The denominator of (\ref{eq:dptdpt_sumrule}) is a result of the different normalization of (\ref{eq:dptdpt}) compared to (\ref{eq:R}), (\ref{eq:C}), and (\ref{eq:D}). To facilitate direct comparison to measured data, we choose not to alter the normalization of \dptdpt. However, (\ref{eq:dptdpt_sumrule}) requires the definition of \dptdpt~to be (\ref{eq:dptdpt}) rather than (\ref{eq:dptdptSTAR}). We show that this change has small effect on measurement in Fig. \ref{fig:Cm_ptSq_pp}. 

Equation (\ref{eq:dptdpt_sumrule}) is equivalent to (\ref{eq:sumrule}) and is a primary result of this paper. Using \eqref{eq:dptdpt_sumrule} we now see, explicitly, the intent of the construction of \dptdpt. First, that it can be interpreted as transverse momentum correlations with multiplicity fluctuations removed $({\cal C}-\langle p_t\rangle^2{\cal R})$, and second that it also removes multiplicity-momentum correlations. Interpreting \dptdpt~in terms of the other observables allows for diagnosis of different physics contributions. For example, simulated results for \R~and \C~are up to an order of magnitude bigger that \D. 
%Any theoretical or phenomenological explanation of that mechanism must address both the cause of correlations and why they are not subtracted by \D~or \R.

In Sec.~\ref{sec:results}, we calculate values of \D~in simulated PYTHIA events. We find that it is on the same order of magnitude as \dptdpt, and possibly even larger. Therefore, \D~should not be assumed negligible when measuring \dptdpt~or \C.

The $1/\langle N\rangle$ (or deviation from $1/\langle N\rangle$) behavior of \dptdpt~can also be studied with (\ref{eq:dptdpt_sumrule}). Notably, \R~exhibits the most obvious expression of the $1/\langle N\rangle$ trend -- see discussion in Sec.~\ref{sec:R} -- and, by construction, \C~and \R~are expected to have similar behavior. This is more obvious in an independent source model, which we discuss in Appendix \ref{sec:ISM}. We test this behavior with simulated events in Sec.~\ref{sec:results}. 

The influence of \apt~also appears in (\ref{eq:sumrule}) and (\ref{eq:dptdpt_sumrule}). Since \apt~is seen to rise with multiplicity, it is a potential source of deviation from $1/\langle N\rangle$ scaling for \dptdpt~that is not due to critical phenomena. 
Since \apt~also increases with increasing collision energy, experiments tested the scaling (\ref{eq:sqrtCmOpt}) for \dptdpt~that shows approximate agreement over a wide range of systems and energies \cite{Adams:2005ka,Adam:2019rsf,Abelev:2014ckr}. The quality of the agreement relies somewhat on the choice of centrality measure. In (\ref{eq:dptdpt_sumrule}), we can see how constituent correlation observables contribute to this scaling and how centrality determination affects this agreement. To avoid interpreting the square root in (\ref{eq:sqrtCmOpt}), we instead consider \dptdpt$/\langle p_t\rangle^2$ and write (\ref{eq:sumrule}) as
\begin{equation}\label{eq:CmOptSq}
\frac{(1+{\cal R})\langle \delta p_{t1} \delta p_{t2}\rangle }{\langle p_{t}\rangle^2}
= \frac{{\cal C}}{\langle p_t\rangle^2}
- \frac{2{\cal D}}{\langle p_t\rangle} 
- {\cal R}.
\end{equation}
Using (\ref{eq:CmOptSq}) we can see that scaling with collision energy requires consistent handling of multiplicity fluctuation \R. Fortunately, if \R~and \C~are measured with the same methods, centrality biases in \C~are subtracted by \R. This is what makes \dptdpt~robust against different centrality definitions.

Alternatively, we can study two-particle transverse momentum correlations by rewriting (\ref{eq:sumrule}), or (\ref{eq:dptdpt_sumrule}), as 
\begin{equation}\label{eq:C_sumrule}
{\cal C} = (1+{\cal R})\langle \delta p_{t1} \delta p_{t2}\rangle  + 2\langle p_t\rangle{\cal D} + \langle p_t\rangle^2{\cal R}.
\end{equation}
Equation (\ref{eq:C_sumrule}) distinguishes the different physical influences on momentum correlations. The rightmost term represents the contribution just from multiplicity fluctuations (including volume fluctuations). This is the largest contribution to \C. 
In this context the quantitative difference between \R~and its momentum weighted counterpart \C~can be measured. \C~is affected by forces like viscosity that impact temperature fluctuations which are represented by the presence of \dptdpt. Similarly, the presence of \D~signals how \C~is influenced by the mechanism that correlates total transverse momentum with multiplicity event-by-event. 

The ALICE collaboration measures the differential quantity $G_2(\Delta \eta, \Delta \phi) = {\cal C}(\Delta \eta, \Delta \phi)/\langle p_t\rangle^2$ \cite{Acharya:2019oxz,Gonzalez:2020gqg,Gonzalez:2020bqm,Gonzalez:2018cty}. Using (\ref{eq:CmOptSq}) we find the integrated version 
\begin{equation}\label{eq:G2}
G_2 = \frac{{\cal C}}{\langle p_t\rangle^2} = 
\frac{(1+{\cal R})\langle \delta p_{t1} \delta p_{t2}\rangle }{\langle p_{t}\rangle^2}
+ \frac{2{\cal D}}{\langle p_t\rangle} 
+ {\cal R},
\end{equation}
but each of the terms on the right hand side can also be measured differentially. For example, the quantities $P_2 = \langle \delta p_{t1}\delta p_{t2}\rangle (\Delta\eta,\Delta\phi)/\langle p_t\rangle^2$ and ${\cal R}(\Delta\eta,\Delta\phi)$ are measured in Ref. \cite{Acharya:2018ddg}. 
With the measurement of ${\cal D}(\Delta\eta,\Delta\phi)/\langle p_t\rangle$, $G_2(\Delta \eta, \Delta \phi)$ can be checked experimentally, using (\ref{eq:G2}). 
 
To summarize, multiplicity fluctuations, \R, set an underlying scale of correlations, (\ref{eq:CorrFn}), that is determined by particle production mechanisms, volume fluctuations, and possibly phase change fluctuations. 
Momentum correlations, \C, indicate both how initial state correlations survive to final state particle \pt, and how transverse momentum can be transferred throughout the collision volume by forces like shear viscosity. 
\D~represents correlations of event-by-event total transverse momentum and multiplicity. Equation (\ref{eq:Dvar}) demonstrates that these correlations are in excess of those from random multiplicity fluctuations, so \D~is therefore tied to particle production. Furthermore, lack of correlations, \D, can signal equilibration while enhancement of \D~could exist around the QGP critical point. 
%Correlations of fluctuations of transverse momentum, \dptdpt, have several theoretical explanations like temperature fluctuations or boosted hot spots. 
%Importantly, the results (\ref{eq:sumrule}), or (\ref{eq:dptdpt_sumrule}), or (\ref{eq:CmOptSq}), or (\ref{eq:C_sumrule}), or (\ref{eq:G2}) suggest that a theoretical or phenomenological explanation of one of the observables \R, \C, \D, or \dptdpt~can be tested by separately addressing each of the others. 
%
%Similarly, with simultaneous experimental measurement of all four observables (\ref{eq:dptdpt}), (\ref{eq:D}) (\ref{eq:R}), and (\ref{eq:C}), Eq. (\ref{eq:sumrule}) acts both as a validation tool for each measurement and as a way to explicitly distinguish multiplicity fluctuations from other correlations mechanisms when looking for critical phenomenon. 

%%%%%%%%%%%%%%%%%%%%%%%%%%%%%%%%%%%%%%%%%%%%%%%%%%%%%%%%%%%%%%%%%%%%%%%%%%%%%%%%%%%%%%%%%%%%%%%
%		RESULTS
%%%%%%%%%%%%%%%%%%%%%%%%%%%%%%%%%%%%%%%%%%%%%%%%%%%%%%%%%%%%%%%%%%%%%%%%%%%%%%%%%%%%%%%%%%%%%%%
\section{Results from Simulation}\label{sec:results}
In this section, our primary goal is to make the first estimates of \D~and test relationship (\ref{eq:sumrule}) with simulated collision events. We do not attempt to preform a comprehensive study using different simulation routines to compare different collision dynamics mechanisms; we leave this for future work. For simplicity, we chose PYTHIA 8.2 \cite{Sjostrand:2014zea} since its description of \pp~collisions is well established and it includes the Angantyr model for nuclear collisions \cite{Bierlich:2018xfw} which provides a baseline estimation based on wounded nucleons. 

%
%\subsection{Multiplicity Dependent Results}\label{sec:res:mult}
%
We look for non-zero values of the new observable \D, defined by equations (\ref{eq:Dvar}) or (\ref{eq:Dint}). This may indicate a deviation from thermal equilibrium, see Sec. \ref{sec:D}. Moreover, we also test the $1/\langle N\rangle$ dependence of \R, \C, \dptdpt, and \D~when using multiplicity as a centrality measure. Deviation from this trend is an indication of non-Poissonian particle production which, in turn, indicates that either particle sources or the particles themselves are not produced independently from event to event, resulting in net correlations. Correlations that develop during the system evolution can also contribute to this behavior.

When measuring correlations based on moments of a multiplicity distribution, centrality biases can be significant especially when the same particles used to calculate the correlations are also used to determine centrality \cite{Luo:2013bmi}. To eliminate centrality biases due to volume fluctuations, 
%-- which set the scale of our correlation observables (see Sec. \ref{sec:R}) -- 
we follow the centrality method of Ref.~\cite{Adam:2019rsf} when calculating observable dependencies on multiplicity. 
This method allows for one-particle-wide multiplicity bins without encountering the effects described at the end of Sec.~\ref{sec:R}. 
%surrounding Eq. (\ref{eq:OneOverN}). 

%
%%%%%%%%%%%%%%%%%%%%%%%%%%%%%%%%%%%%%% FIG: <<N>> %%%%%%%%%%%%%%%%%%%%%%%%%%%%%%
\begin{figure}
	%\centering
	\includegraphics[width=\linewidth]{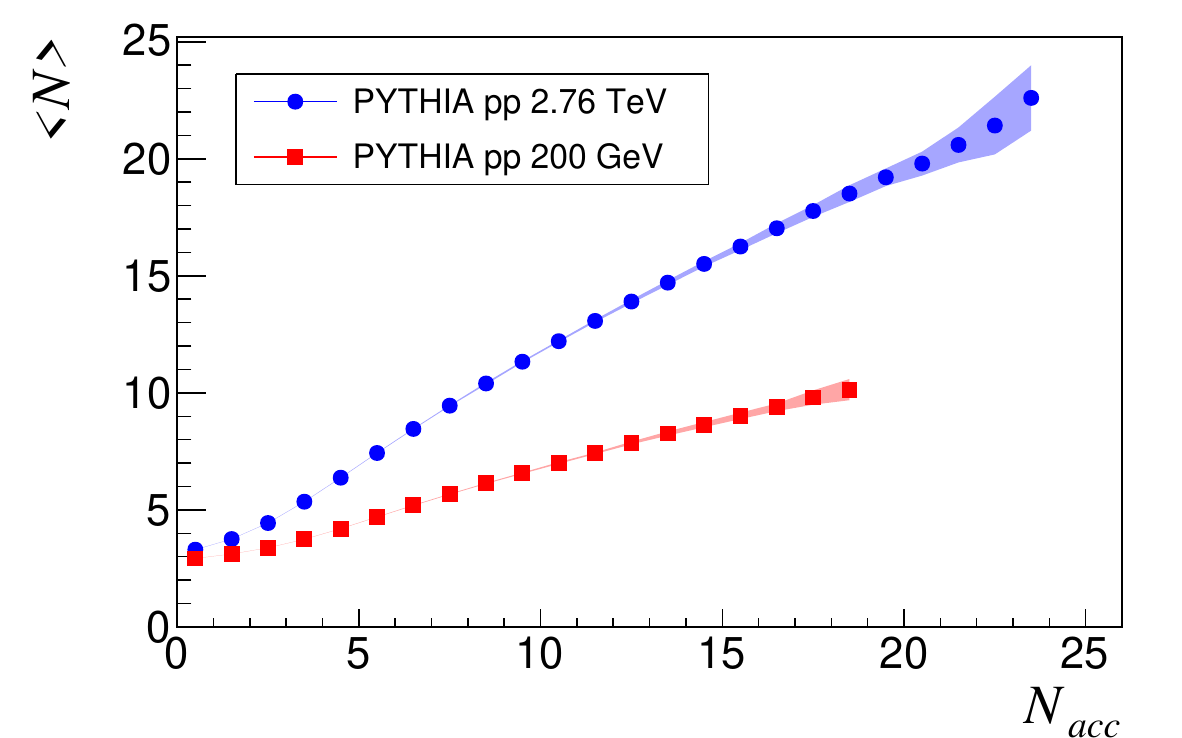}
	\caption{Sub-group averaged mid-rapidity multiplicity $\langle N\rangle$ as a function of accepted multiplicity $N_{acc}$ in the region $0.5 < |\eta| < 1.0$ for \pp~200 GeV and $0.5 < |\eta| < 0.8$ for \pp~2.76 TeV. }
	\label{fig:N_pp}
	%\centering
	\includegraphics[width=\linewidth]{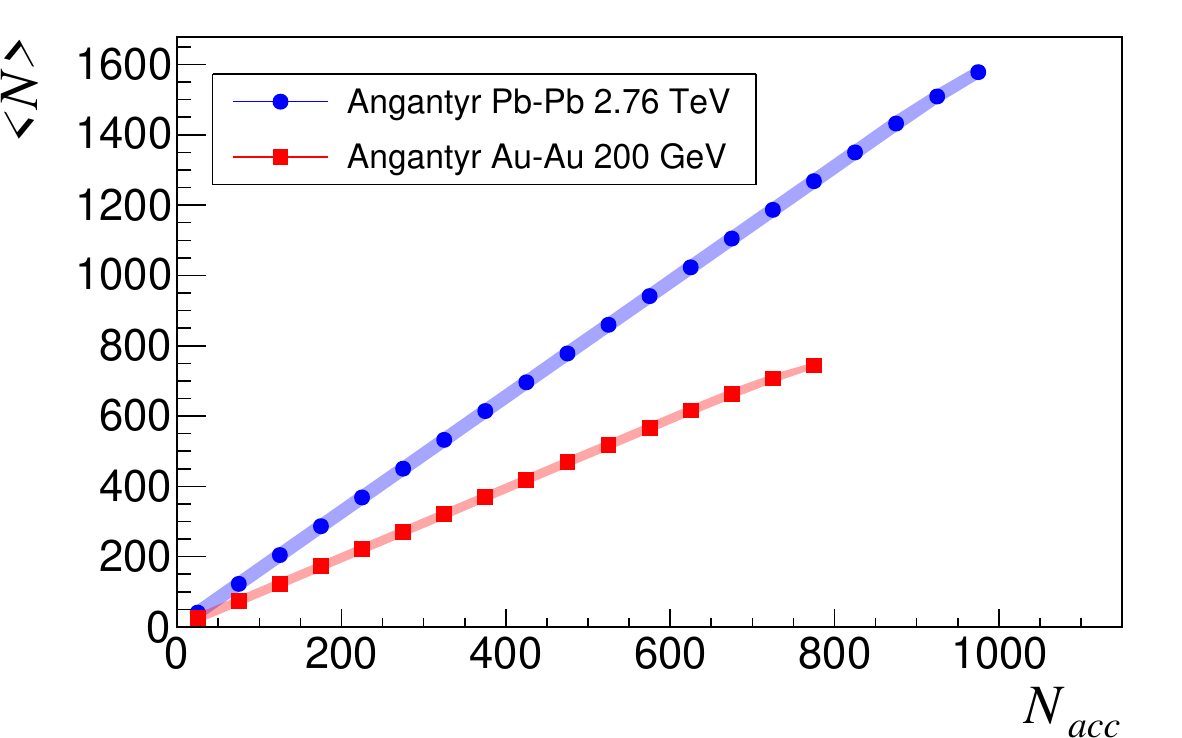}
	\caption{Sub-group averaged mid-rapidity multiplicity $\langle N\rangle$ as a function of accepted multiplcity $N_{acc}$ in the region $0.5 < |\eta| < 1.0$ for Au-Au 200 GeV and $0.5 < |\eta| < 0.8$ for Pb-Pb 2.76 TeV.}
	\label{fig:N_AA}
\end{figure}

In this method, observables are calculated using all charged particles in the mid-rapidity region $|\eta|<0.5$ while centrality is determined using all charged particles in the remaining region of experimental rapidity acceptance. We label these accepted centrality determining particles $N_{acc}$. For comparison to STAR, charged particles in the region $0.5 < |\eta| < 1.0$ are used for $N_{acc}$. For comparison to ALICE, charged particles in the region $0.5 < |\eta| < 0.8$ are used for $N_{acc}$. 
In Figs.~\ref{fig:N_pp} and \ref{fig:N_AA} we plot the average mid-rapidity multiplicity vs $N_{acc}$ in PYTHIA events. 
The acceptance difference between STAR and ALICE accounts for the different slopes in the mid-rapidity multiplicities. 
This centrality measure also has the consequence of transforming the two-particle correlation observables into three-particle correlations since two particles are used to calculate the correlation and different particles are used to determine $N_{acc}$. 
If the pseudorapidity distribution of charged particles is approximately flat in the rapidity acceptance, then the correlation between the number of particles in the centrality determining and mid-rapidity regions is effectively 1, and multiplicity trends can be taken at face value. 

A nonlinear correlation between $N_{acc}$ and the mid-rapidity multiplicity could induce some modulation in correlation measurements away from the expected  $1/\langle N\rangle$ trend. 
However, the average mid-rapidity multiplicity $\langle N \rangle$ tracks very linearly with $N_{acc}$ for PYTHIA events in Figs.~\ref{fig:N_pp} and \ref{fig:N_AA} for \pp~and $AA$ collisions respectively.

Similarly to Ref.~\cite{Adam:2019rsf}, we employ the so-called ``sub-group'' method for estimating the uncertainty of correlation observables. In our analysis, the full set of events for a given centrality class is divided up into 30 sub-groups and all observables are calculated for each sub-group. Each observable is then averaged over all sub-groups and the standard deviation is used to estimate the uncertainty. 
For $AA$ collisions, when multiplicity is used for centrality and after taking the sub-group average, we average observable values over several multiplicity bins and the set the error band to represent the standard deviation of those values. 

In Figs.~\ref{fig:avgpt_pp} and \ref{fig:avgpt_AA} we report the average transverse momentum per particle for all charged particles from PYTHIA events in select \pp~and $AA$ systems and energies. For all PYTHIA simulation results we show in this work, we have used both the centrality and sub-group methods described above. 
Notice the increase in average \pt~per particle as multiplicity increases in both figures. We argue in Sec.~\ref{sec:D} that this is important for understanding multiplicity-momentum correlations \D. 
The smaller increase in $AA$ collisions compared to \pp~collisions is likely a factor in the different magnitudes of \D~estimates from different collision systems. 
We also show STAR data for negative pions, from Ref. \cite{Abelev:2008ab}, to illustrate that experimental measurements also find a comparable increase in \apt~with multiplicity. The overall difference in magnitude of the STAR data compared to our PYTHIA calculation is due to the fact that the PYTHIA calculation includes all charged particles. 

We now turn to estimating observables \R, \C, \dptdpt, and \D, 
and their mathematical relationship (\ref{eq:sumrule}) with PYTHIA/Angantyr simulations of \pp~and $AA$ collision systems at select energies. 
For the four observables we analyze events following  (\ref{eq:R}), (\ref{eq:C}), (\ref{eq:dptdpt}), and (\ref{eq:D}) respectively. We use charged particles in the transverse momentum range $0.15 < p_t < 2~GeV$ for both Au-Au collisions at  $\sqrt{s}=200~GeV$ and for Pb-Pb collisions at  $\sqrt{s}=2.76~TeV$. To identify deviation from $1/\langle N\rangle$ behavior we plot the product of each observable with the multiplicity $\langle N\rangle$. If there is no deviation from $1/\langle N\rangle$, then results will be constant with multiplicity, though with different magnitudes. 

\begin{figure}
	%\centering
	\includegraphics[width=\linewidth]{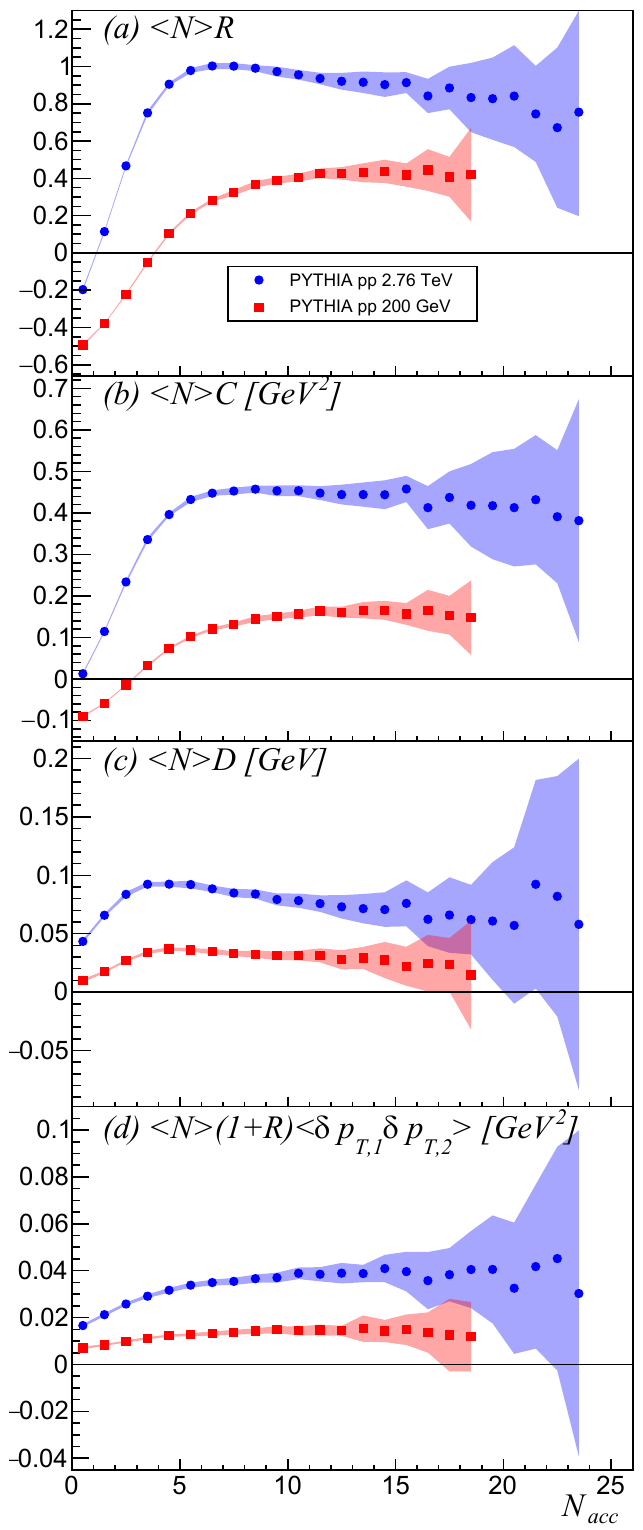}%
	\caption{\label{fig:combine_pp} Calculation of observables \eqref{eq:Rint}, (\ref{eq:Cint}), (\ref{eq:dptdptcorr}), and (\ref{eq:Dint}), scaled by mid-rapidity multiplicity $\langle N\rangle$ using PYTHIA $pp$ collisions. }
\end{figure}

%Multiplicity fluctuations, attributed to volume fluctuations, \R, are defined by (\ref{eq:Rint}) or (\ref{eq:R}). 
Results for \NR~from PYTHIA simulation of \pp~collisions at $200~GeV$ and $2.76~TeV$ are shown in Fig. \ref{fig:combine_pp}(a). At lower multiplicities, the deviation from $1/\langle N\rangle$ behavior and the transition to negative values are both the result of a small variance of \textit{total} multiplicity produced in these events. Consider that events with very few particles in the centrality defining rapidity region also have correspondingly very few particles in the mid-rapidity region. In this case the variance of mid-rapidity particles is nearly zero. Following the argument surrounding Eq.~(\ref{eq:OneOverN}), negative values of \R~can be expected. At larger multiplicities \NR~becomes more flat and the error band increases with the scarcity of events.

It is interesting to note that \NR~in \pp~collisions at $\sqrt{s}=2.76~TeV$ indicates a slightly faster than $1/\langle N\rangle$ decrease with increasing multiplicity when compared to $\sqrt{s}=200~GeV$ collisions. It will be interesting to discover if this change persists to higher or lower collision energies in both simulation and experiment. Moreover, it is also significant to point out that \NR~is non-zero. This indicates that particle production -- averaged over events -- is not Poissonian and therefore not independent. This reinforces the fact that \R~measures a fundamental particle production mechanism. Deviation of experimental measurements from PYTHIA estimates could signal the contribution form different particle sources. A comparison covering different collision systems and energies may be a useful tool to characterize the onset of QGP or jet influences on particle production.

\begin{figure}
	%\centering
	\includegraphics[width=\linewidth]{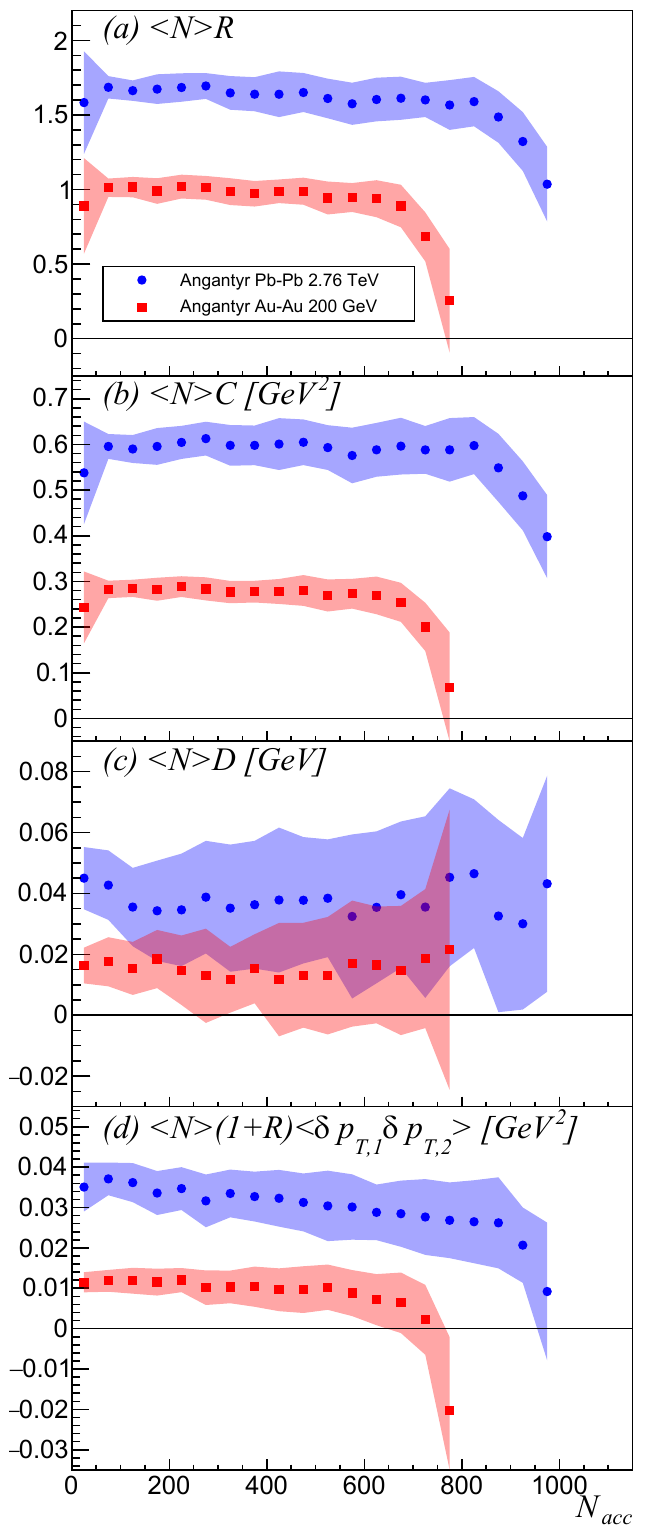}%
	\caption{\label{fig:combine_AA} Calculation of observables (\ref{eq:Rint}), (\ref{eq:Cint}), (\ref{eq:dptdptcorr}), and (\ref{eq:Dint}), scaled by mid-rapidity multiplicity $\langle N\rangle$ using PYTHIA/Angantyr $AA$ collisions. }
\end{figure}

Results for \NR~from PYTHIA/Angantyr simulation of Au-Au and Pb-Pb collisions at $\sqrt{s}=200~GeV$ and $\sqrt{s}=2.76~TeV$ are shown in Fig.~\ref{fig:combine_AA}(a), plotted versus multiplicity, $N_{acc}$. When centrality is determined by multiplicity, \NR~is seemingly constant until the most central points. The deviation in high multiplicity events is likely due to low statistics. 
The drop of the lowest multiplicity point is the result of averaging the first few lowest multiplicity bins where values may be small or negative for the same reasons small or negative values appeared in low multiplicity \pp~collisions. In general, the approximately constant value of \NR~with multiplicity is consistent with a superposition of \pp~sub-collision model. 

\begin{table}
	\centering
	\caption{\label{tab:obs_pp} List of integrated values of observables $\langle N\rangle_{pp}$, \apt$_{pp}$,  (\ref{eq:R}), (\ref{eq:C}), (\ref{eq:dptdpt}), and (\ref{eq:D}) using PYTHIA \pp~collision events calculated with the sub-group method. Calculations are made with charged particles from the kinematic region and $|\eta|<0.8~(\sqrt{s}=2.76~TeV)$ or $|\eta|<1.0~(\sqrt{s}=200~GeV)$. Listed uncertainties are the standard deviation of the sub-group values.}
	
	\renewcommand{\arraystretch}{1.2} % make text lines thicker so text doesn't touch \hline
	\begin{tabular}{|c|c|c|c|c|}
		\hline
		$\sqrt{s}$				& $200~GeV$	&$\pm$	& $2.76~TeV$	& $\pm$ \\
		\hline
		$\langle N\rangle_{pp}$	& 6.635		& 3.65$\times 10^{-3}$	& 8.453		& 8.10$\times 10^{-3}$ \\
		\hline
		\apt$_{pp}$				& 0.4860 	& 1.33$\times 10^{-4}$	& 0.5356	& 1.78$\times 10^{-4}$ \\
		\hline
		${\cal R}_{pp}$			& 0.2731	& 7.58$\times 10^{-4}$	& 0.453		& 1.02$\times 10^{-3}$ \\
		\hline
		${\cal C}_{pp}$			& 0.0842	& 2.20$\times 10^{-4}$	& 0.1738	& 4.84$\times 10^{-4}$ \\
		\hline
		\dptdpt$_{pp}$			& 0.00257	& 2.27$\times 10^{-5}$	& 0.00446	& 3.67$\times 10^{-5}$ \\
		\hline
		${\cal D}_{pp}$			& 0.01685	& 9.32$\times 10^{-5}$	& 0.0348	& 1.68$\times 10^{-4}$ \\
		\hline
	\end{tabular}
\end{table}
%
%

%
%%%%%%%%%%%%%%%%%%%%%%%%%%%%%%%%%%%%%% FIG: combined pp %%%%%%%%%%%%%%%%%%%%%%%%%%%%%%
%
%

%
%%%%%%%%%%%%%%%%%%%%%%%%%%%%%%%%%%%%%% FIG: combined AA %%%%%%%%%%%%%%%%%%%%%%%%%%%%%%

In Appendix~\ref{sec:ISM} we calculate \R~and the other correlation observables using an Independent Source Model (ISM). In this model each event is comprised of $K$ independent random sources where $K$ fluctuates form event to event. Each source also yields a random number of particles that fluctuates form source to source. We then choose wounded nucleon sources as a test case and estimate the average correlation of a single source using PYTHIA \pp~collisions. Values for \pp~quantities are listed in Table~\ref{tab:obs_pp}.

%
%%%%%%%%%%%%%%%%%%%%%%%%%%%%%%%%%%%%%% FIG: combined Np %%%%%%%%%%%%%%%%%%%%%%%%%%%%%%
\begin{figure}
	%\centering
	%\vspace{-3mm}
	\includegraphics[width=\linewidth]{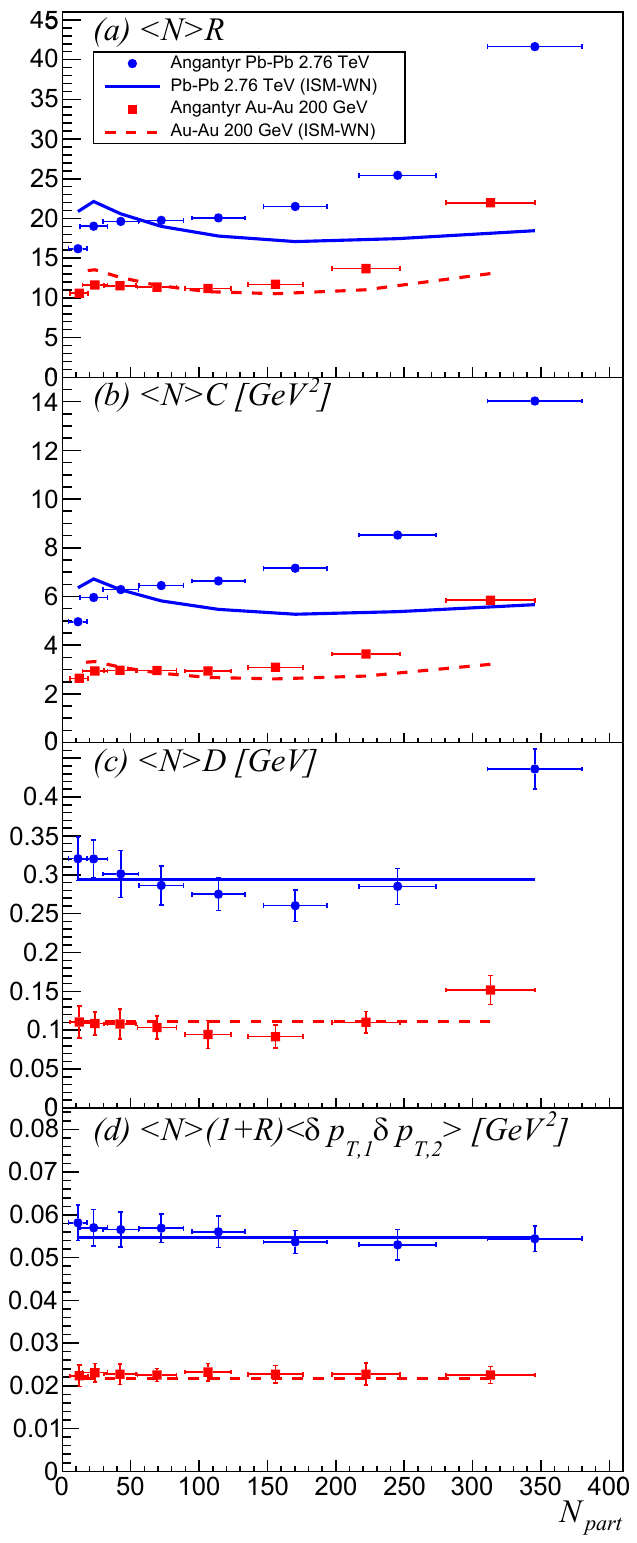}%
	\caption{\label{fig:combine_Np} Calculation of observables 
		(\ref{eq:Rint}) - (\ref{eq:dptdptcorr}), 
		%		(\ref{eq:Rint}), (\ref{eq:Cint}), (\ref{eq:Dint}), and (\ref{eq:dptdptcorr}), 
		scaled by multiplicity $\langle N\rangle$ using PYTHIA/Angantyr $AA$ collisions. Centrality is determined following experimental percent centrality scheme \cite{STAR:2008med, ALICE:2013hur}. Solid and dashed lines represent the wounded nucleon model using values from Table \ref{tab:obs_pp} and taking the average and variance of participants from PYTHIA/Angantyr.}
\end{figure}

In this work we attribute all volume fluctuations to source number fluctuations. To see how multiplicity fluctuations are influenced by volume (source) fluctuations, imagine that participant sources are independent, then the variance of the number of participants in the numerator of the rightmost term of (\ref{eq:R_WN}) follows Poisson statistics. In that case,  $Var(N_{part})=\langle N_{part}\rangle$, then Eq.~(\ref{eq:R_WN}) becomes ${\cal R}=(2{\cal R}_{pp}+1)/\langle N_{part}\rangle$. Note that the contribution from source correlations is represented by $2{\cal R}_{pp}$. If $2{\cal R}_{pp}=1$, then half of multiplicity fluctuations come from source correlations and half come from volume fluctuations. If $2{\cal R}_{pp} < 1$, then volume fluctuations contribute more to \R~than source correlations. If $2{\cal R}_{pp} > 1$, then volume fluctuations contribute less to \R~than source correlations. We estimate ${\cal R}_{pp}$ using PYTHIA simulations and list values in Table \ref{tab:obs_pp} for $\sqrt{s}=200~GeV$ and $\sqrt{s}=2.76~TeV$ collision energies. At $\sqrt{s}=200~GeV$, a bit less than two thirds of \R~comes from volume fluctuations. At $\sqrt{s}=2.76~TeV$, about half of \R~comes from volume fluctuations.

To compare to the ISM, in Fig.~\ref{fig:combine_Np}(a) we plot PYTHIA/Angantyr results for \NR~with respect to the number of participating nucleons $(N_{part})$ corresponding to percent centrality event ensembles. The most central events are the top 10\% highest multiplicity events, and each subsequently peripheral point uses the next 10\% of the remaining highest multiplicity events. 
The solid and dashed lines represent the ISM with wounded nucleon sources, ISM-WN, from (\ref{eq:R_WN}). The average and variance of participants are calculated from PYTHIA/Angantyr in corresponding percent centrality ranges.  
%
%\red{This assumption has the advantage that the calculation of \NR~is robust against binning of events by participants.} 
%Also using PYTHIA, we calculate the ``integrated'' value ${\cal R}_{pp}$ including all \pp~events without centrality constraints.  
%Values for \pp~collisions at $\sqrt{s}=200~GeV$ and $\sqrt{s}=2.76~TeV$ are listed in Table \ref{tab:obs_pp}. 
%When defining centrality with $N_{part}$, we use the full experimental rapidity acceptance to calculate all observables and the same is done for integrated values. 

The variance of participants is extremely sensitive to the centrality definition, so we choose to use standard experimental definitions. This sensitivity is the cause of the nonlinear shape of the ISM-WN  curves in Fig.~\ref{fig:combine_Np}. For purely independent sources, the distributions of participants is Posisson so that the variance of participants equals the mean in Eq.~(\ref{eq:R_WN}). In that situation, ISM curves on Fig.~\ref{fig:combine_Np}(a) would take constant values of \NR$_{200GeV}=5.130$ and \NR$_{2.76TeV}=8.056$. Differences between ISM-WN and PYTHIA results are likely due in part to the particle production mechanism and in part to multi-particle interactions. Notice that the deviation form ISM-WN is largest in the most central collisions where greatest variety of events is present. 

Note that in Fig.~\ref{fig:combine_Np}, we report PYTHIA/Angantyr results with horizontal and vertical error bars. 
%In figures where multiplicity is used to estimate centrality, an error band indicates the uncertainty of sub-group observable values given a fixed multiplicity. 
%In Figs.~\ref{fig:Cm_ptSq_Np} and \ref{fig:combine_Np}, 
When nucleon participants are used to estimate centrality, uncertainty arises from both fluctuations in participant number as well as fluctuations in the observable values. Since centrality is not determined directly by multiplicity, all charged particles in the experimental acceptance can be used to calculate observables. Consequentially, the fluctuations of calculated observable values is reduced, but this improvement is exchanged for significant uncertainty in the determination of the number of participants. This is represented by the horizontal error bars in Figs.~\ref{fig:Cm_ptSq_Np} and \ref{fig:combine_Np}. 

%Notice that when $N_{part}=2$ on Fig.~\ref{fig:combine_Np}(a), Eq.~(\ref{eq:R_WN}) matches closely with the data, but deviates at larger $N_{part}$. This may indicate that the source value ${\cal R}_{s} = 2{\cal R}_{pp}$ is dominated by lower multiplicity \pp~events, or simply that participating nucleons, on their own, are not a good indicator of all particle sources. 
%
%\red{To test this, we relax the assumption that sources are purely independent and calculate the variance of participants in PYTHIA/Angantyr for use in (\ref{eq:R_WN}). The variance of participants is extremely sensitive to centrality binning, so we choose to calculate the average and variance of participants based on experimentally defined percent centrality binning. }

%Momentum correlations, \C, are defined by equations (\ref{eq:Cint}) or (\ref{eq:C}). 
Due to its similar construction, \C~shares many of the same centrality trends as \R. \C~scales both with \R~and \apt$^{2}$; the latter scaling is visible by examining the ISM Eq.~(\ref{eq:C_WN}). 
%For a Poissonian distribution of participant sources we have ${\cal C}=(2{\cal C}_{pp}+\langle p_t\rangle^2)/\langle N_{part}\rangle$.  
%Source correlations and volume fluctuations have equal contributions to \C~when $2{\cal C}_{pp}=\langle p_t\rangle^2$. Using values from Table \ref{tab:obs_pp} we find that at the contributions to \C~from source correlations and volume fluctuations are roughly equal but volume fluctuations are slightly larger at $\sqrt{s}=200~GeV$ and source correlations are slightly larger at $\sqrt{s}=2.76~TeV$. We leave an analysis of collision energy dependence to future work.
%
\NC~is reported on Figs.~\ref{fig:combine_pp}(b), \ref{fig:combine_AA}(b), and \ref{fig:combine_Np}(b). Centrality behaviors mostly follow those of \NR~for both PYTHIA and ISM-WN results. 
%except in Fig.~\ref{fig:combine_Np}(b). In comparison to ISM for wounded nucleons in Pb-Pb collisions at $\sqrt{s}=2.76~TeV$, \NC~increases from peripheral collisions peaking around $N_{part}\approx 100$. This increase is not seen in \NR~for the same collision system. Additionally this peak is not seen for \NC~in Au-Au collisions at $\sqrt{s}=200~GeV$. 
For purely independent sources, ISM results take constant values of \NC$_{200GeV}=1.342~GeV^2$ and \NC$_{2.76TeV}=2.682~GeV^2$ on Fig.~\ref{fig:combine_Np}(b). 

%The centrality dependence of \C~can be analyzed in the context of (\ref{eq:C_sumrule}). 
%Notice in Fig.~\ref{fig:combine_Np}(b) that \ND~and \Ndptdpt~both exceed the wounded nucleon model expectation in the same region where \NC~peaks. Care should be taken when assigning meaning to the peak in Fig.~\ref{fig:combine_Np}(b): when \NC~is plotted with respect to multiplicity in Fig.~\ref{fig:combine_AA}(b), the peak behavior is not seen. Therefore, we use these observations only as a tool to demonstrate the usefulness of measuring correlation observables as a complementary set with a mathematical connection like (\ref{eq:sumrule}). 

%Correlations of transverse momentum fluctuations, \dptdpt, are defined by equations (\ref{eq:dptdptcorr}) or (\ref{eq:dptdpt}). STAR and ALICE have measured the similar form (\ref{eq:dptdptSTAR}), usually reporting it as (\ref{eq:sqrtCmOpt}). 
In Fig.~\ref{fig:Cm_ptSq_pp} we compare \dptdpt~as Eq.~(\ref{eq:sqrtCmOpt}), calculated from both (\ref{eq:dptdpt}) and (\ref{eq:dptdptSTAR}), with experimental data for \pp~collisions. The two methods (\ref{eq:dptdpt}) and (\ref{eq:dptdptSTAR}) are in generally good agreement. Similar results are obtained for $AA$ collisions but are omitted from Fig. \ref{fig:Cm_ptSq_AA} for clarity. 

PYTHIA/Angantyr comparisons to experimental $AA$ data are shown in Fig.~\ref{fig:Cm_ptSq_AA}. Agreement with STAR data is good but significant difference from ALICE data is seen. This is likely due to differences in the multiplicity centrality determination between STAR and ALICE; our calculation technique reproduces that of STAR.  

To test the $1/\langle N\rangle$ dependence of \dptdpt, we plot \Ndptdpt~in Figs. \ref{fig:combine_pp}(d), \ref{fig:combine_AA}(d), and \ref{fig:combine_Np}(d). As we discover in Sec.~\ref{sec:sumrule} and Appendix \ref{sec:ISM}, the factor $(1+{\cal R})$ is required to re-scale the normalization of (\ref{eq:dptdpt}) so that it follows the same $1/\langle N\rangle$ trend as the other observables. 

PYTHIA results for \Ndptdpt~in \pp~collisions, shown in Fig.~\ref{fig:combine_pp}(d), are mostly flat except in peripheral collisions where fluctuations become small. In $AA$ collisions, shown in Fig.~\ref{fig:combine_AA}(d), the trend is again consistent, inside the error band, with a $1/\langle N\rangle$ except in the most central collisions that are statistically limited. 
\dptdpt~has reduced effect from small fluctuations at low multiplicity in comparison to \R~or \C. By construction, \dptdpt~removes multiplicity fluctuations (see the discussion following Eq.~(\ref{eq:dptdpt_sumrule}) in Sec.~\ref{sec:sumrule}). 
Therefore, \dptdpt~appears insensitive to choice of centrality via multiplicity or participating nucleons. Results in Fig.~\ref{fig:combine_Np}(d) for \Ndptdpt~with respect to participating nucleons are also constant and in strong agreement with the wounded nucleon model, Eq.~(\ref{eq:dptdpt_WN}). 

The deviation of the wounded nucleon model in Fig.~\ref{fig:Cm_ptSq_Np} in comparison to Fig.~\ref{fig:combine_Np}(d) is likely due to multiple factors. First, the integrated value of \apt$_{pp}$ is used with (\ref{eq:dptdpt_WN}). In the independent source model, \apt~is the same for individual sources as it is for the whole event. 
\apt$_{pp}$ does not change value in our simple wounded nucleon model, but \apt~does change in the centrality dependent measurement. 
Last, the factor of $(1+{\cal R})$ in the denominator of (\ref{eq:dptdpt_WN}) also induces a difference from PYTHIA values. As we see in Fig.~\ref{fig:combine_Np}(a), \R~for our wounded nucleon model is generally smaller than PYTHIA values, particularly in more central collisions. 

%Multiplicity-momentum correlations, \D, are defined by equations (\ref{eq:Dvar}) or (\ref{eq:Dint}). 
An objective of this work is to stimulate experimental measurement of \D. The first estimates of \ND~from PYTHIA/Angantyr \pp~and $AA$ collisions are shown in Figs.~\ref{fig:combine_pp}(c), \ref{fig:combine_AA}(c), and \ref{fig:combine_Np}(c). Immediate observations include that ${\cal D}\neq 0$ and is positive. 
The positive nonzero value of \D~is consistent with \apt~centrality trends. For example, notice in Fig.~\ref{fig:avgpt_pp} that the average transverse momentum per particle increases with the number of particles. This is a multiplicity-momentum correlation. The difference of magnitudes of \ND~in \pp~and $AA$ collisions may be due to the fact that the rate of increase of \apt~with multiplicity is greater in \pp~collisions that in $AA$ for PYTHIA simulations.

The flatness of \ND~with respect to multiplicity in Fig.~\ref{fig:combine_AA}(c) indicates agreement with the $1/\langle N\rangle$ dependence.  
%Deviation from a $1/\langle N\rangle$ trend is expected at low multiplicities for the same reasons as stated for \R. 
Interestingly, \pp~collisions in  Fig.~\ref{fig:combine_pp}(c) show a small negative slope with increasing multiplicity, indicating a faster than $1/\langle N\rangle$ drop with increasing multiplicity. This slope seems to increase from $\sqrt{s}=200~GeV$ to $\sqrt{s}=2.76~TeV$ collision energies. We look for experimental measurements in a larger range of collision energies to examine this behavior. 

Figure \ref{fig:combine_Np}(c) shows that when centrality is determined by percent centrality, \D~values from PYTHIA/Angantyr are in the general range of our wounded nucleon model. Deviations may simply signal a difference between our choice to use only participant nucleons as sources in our independent source model and the PYTHIA/Angantyr model. Larger deviations of the most central points may indicate increased importance of multiplicity-momentum fluctuations when interpreting momentum correlation observables like \C, $G_2$, and \dptdpt; the contributions of multiplicity-momentum correlations to those observables is discussed in Sec.~\ref{sec:sumrule} around equations (\ref{eq:dptdpt_sumrule}), (\ref{eq:C_sumrule}), and (\ref{eq:G2}).

\D~may also be sensitive to the thermalization of the medium. In Sec.~\ref{sec:D} we calculate that \D~vanishes in equilibrium in in the Grand Canonical Ensemble. Therefore, non-zero measurements of \D~indicate incomplete thermalization. Furthermore, we show in Ref.~\cite{Gavin:2016nir} that \dptdpt~can be used to quantify incomplete thermalization. This suggests that \D~and \C~may be used in the same way to provide additional constraints on that model. We leave this to future work.

Finally, as a validation, we calculate Eq.~(\ref{eq:sumrule}) for every point in Figs.~\ref{fig:combine_pp}, \ref{fig:combine_AA}, and \ref{fig:combine_Np}. We find minimal numerical error with agreement to the level of  $|3|\times 10^{-17}$ or smaller for all points. 

%%%%%%%%%%%%%%%%%%%%%%%%%%%%%%%%%%%%%%%%%%%%%%%%%%%%%%%%%%%%%%%%%%%%%%%%%%%%%%%%%%%%%%%%%%%%%%%
%		Summary
%%%%%%%%%%%%%%%%%%%%%%%%%%%%%%%%%%%%%%%%%%%%%%%%%%%%%%%%%%%%%%%%%%%%%%%%%%%%%%%%%%%%%%%%%%%%%%%
\section{Summary}\label{sec:summary}
%
%In Sec.~\ref{sec:correlations}, we briefly discuss the construction of a general two-particle momentum density correlation function, Eq.~(\ref{eq:CorrFn}). 
We discuss four two-particle correlation observables: multiplicity fluctuations, \R, Eq.~(\ref{eq:Rint}), transverse momentum correlations, \C, Eq.~(\ref{eq:Cint}), net correlation of transverse momentum fluctuations, \dptdpt, Eq.~(\ref{eq:dptdptcorr}), and multiplicity-momentum correlations \D, Eq.~(\ref{eq:Dint}). Importantly, all of these observables are derived from this same common origin, Eq.~(\ref{eq:CorrFn}) and we find an observable mathematical connection between them, Eq.~(\ref{eq:sumrule}). 
We estimate these observables and their connection with PYTHIA/Angantyr simulated collisions events at $\sqrt{s}=200~GeV$ and $\sqrt{s}=2.76~TeV$ collision energies. 

Multiplicity-momentum correlations are a new observable estimated here for the first time. We propose that this collection of observables and their mathematical relationship, measured or calculated simultaneously, can provide more information than separately examining the individual observables. Measurements of these observables over a wide range of collision systems and energies may provide invaluable information about initial state particle production mechanisms of hadronic collisions as well as observable influences of local equilibration on two-particle correlation measurements. 

\begin{acknowledgments}
This work is supported in part by NSF-PHY1913005 (G.M. and M.K.). G.M. would like to give special thanks to undergraduates Alec Ferensic, Reinali Calisin, and Neil Fraylick for assistance in this work.
\end{acknowledgments}

\appendix
%%%%%%%%%%%%%%%%%%%%%%%%%%%%%%%%%%%%%%%%%%%%%%%%%%%%%%%%%%%%%%%%%%%%%%%%%%%%%%%%%%%%%%%%%%%%%%%
%		INDEPENDENT SOURCE MODEL
%%%%%%%%%%%%%%%%%%%%%%%%%%%%%%%%%%%%%%%%%%%%%%%%%%%%%%%%%%%%%%%%%%%%%%%%%%%%%%%%%%%%%%%%%%%%%%%
\section{Independent Source Model}\label{sec:ISM}
The independent source model assumes that nuclear collision events are comprised of a superposition of independent sources of particles and ignores any interactions between particles emitted from different sources. Each event has a fluctuating number of sources and each source has a fluctuating multiplicity and momentum distribution of particles. In this section we detail how the observables discussed in sections \ref{sec:R}, \ref{sec:C}, \ref{sec:dptdpt}, and  \ref{sec:D},  depend on both of these types of fluctuation. A similar discussion for only \R~and \dptdpt~appears in Ref. \cite{Gavin:2011gr}. 

Our independent source model assumes that a single collision event is the sum of $K$ independent particle sources. Each source is represented by a momentum distribution $\hat{\rho}_1(\mathbf{p})$ normalized such that $\int \hat{\rho}_1(\mathbf{p}) d^3\mathbf{p}=\mu$ is the mean multiplicity per source.
To understand the average particle distribution of sources, imagine a large number of sources running from $k=1,..., N_{src}$, where each emits $n_k$ particles. 
The average number of particles per source is then
\begin{equation}\label{eq:src_navg}
	\bar{n}=\frac{1}{N_{src}}\sum_{k=1}^{N_{src}}\sum_{i=1}^{n_k}1\xrightarrow[N_{src}\rightarrow\infty]{}\int \hat{\rho}_1(\mathbf{p}) d^3\mathbf{p}=\mu,
\end{equation}
where the overbar indicates an average over sources, and $\hat{\rho}_1(\mathbf{p})$ is the particle momentum distribution per source in the limit of a continuum of all possible sources. In that limit, each source multiplicity has mean $\mu=\bar{n}$ and variance $\sigma^2=\overline{n^2}-\bar{n}^2$. Similarly, the distribution of particle pairs emitted from one source is
\begin{eqnarray}\label{eq:src_pairs}
	\overline{n(n-1)} &=& \frac{1}{N_{src}}\sum_{k=1}^{N_{src}}\sum_{i=1}^{n_k}\sum_{j\neq i}^{n_k}1  \\ \nonumber
	&\xrightarrow[N_{src}\rightarrow\infty]{}& \iint \hat{\rho}_2(\mathbf{p}_1,\mathbf{p}_2) d^3\mathbf{p}_1 d^3\mathbf{p}_2 = \sigma^2 + \mu^2 - \mu,
\end{eqnarray}
where $\hat{\rho}_2(\mathbf{p}_1,\mathbf{p}_2)$ is the particle pair momentum distribution for an individual source.

The event averaged singles and pair momentum distributions become
\begin{eqnarray}
	\rho_1 &=& \left\langle\, \hat{\rho}_1(\mathbf{p}) K\,  \right\rangle \label{eq:evnt_Ksrc_singles} \\
	&\text{and}& \nonumber \\
	\rho_2 &=& \left\langle\, \hat{\rho}_2(\mathbf{p}_1,\mathbf{p}_2) K + \hat{\rho}_1(\mathbf{p}_1) \hat{\rho}_1(\mathbf{p}_2) K (K-1)\, \right\rangle, \label{eq:evnt_Ksrc_pairs}
\end{eqnarray}
where the angled brackets indicate the average over events and each event has $K$ independent sources. 
Equation (\ref{eq:evnt_Ksrc_singles}) specifies that the event multiplicity is a superposition of $K$ sources, yielding
\begin{equation}\label{eq:avgN_src}
	\langle N\rangle = \langle K\rangle \mu. 
\end{equation}
Equation (\ref{eq:evnt_Ksrc_pairs}) indicates particle pairs are made up of the sum of pairs from the $K$ individual sources, each with $\hat{\rho}_2$ pairs, plus the sum of pairs where one particle of the pair is from one source and the other particle comes from a different source. For one pair of sources the particle pair distribution is $\hat{\rho}_1 \hat{\rho}_1$, and there are $K(K-1)$ pairs of sources. 
%Since $\hat{\rho}_1$ and $\hat{\rho}_2$ are the same for each independent source, they can factor out of the event average where appropriate.
The event average number of particle pairs then becomes 
\begin{equation}\label{eq:avgPairs_src}
	\langle N(N-1)\rangle = \langle K\rangle (\sigma^2-\mu)+\langle K^2\rangle\mu^2.
\end{equation}

Beginning with \R~as defined in (\ref{eq:Rint}) with (\ref{eq:CorrFn}), then using (\ref{eq:evnt_Ksrc_singles}) and (\ref{eq:evnt_Ksrc_pairs}), we find
\begin{equation}\label{eq:R_ISM}
	{\cal R} = 
	\frac{{\cal R}_s}{\langle K\rangle}
	%\left(\frac{\sigma^2 - \mu}{\mu^2}\right)\frac{1}{\langle K\rangle} 
	+\frac{\langle K^2\rangle - \langle K\rangle^2}{\langle K\rangle^2}
\end{equation}
where ${\cal R}_s=(\sigma^2 - \mu)/\mu^2$ is the equivalent of (\ref{eq:R}) for sources when averaging is done over the ensemble of all possible independent sources. Event-by-event fluctuations in the number of sources are characterized by the variance of $K$ in the rightmost term. Since the sources are taken to be independent, this variance follows Poisson statistics, so $\langle K^2\rangle - \langle K\rangle^2 = \langle K\rangle$, and therefore fluctuations (\ref{eq:R_ISM}) are diminished by $\langle K\rangle^{-1}$.

Two-particle correlations of transverse momentum, \C, are defined by Eq.~(\ref{eq:Cint}). Using Eq.~(\ref{eq:CorrFn}) with (\ref{eq:evnt_Ksrc_singles}) and (\ref{eq:evnt_Ksrc_pairs}) we find
\begin{equation}\label{eq:C_ISM}
	{\cal C} = 
	\frac{{\cal C}_s}{\langle K\rangle}
	%\left(\frac{\sigma^2_{P_T} - \mu\langle p_t\rangle}{\mu^2}\right)\frac{1}{\langle K\rangle} 
	+\left(\frac{\langle K^2\rangle - \langle K\rangle^2}{\langle K\rangle^2}\right)\langle p_t\rangle^2,
\end{equation}
where ${\cal C}_s=(\sigma^2_{P_T} - \mu\langle p_t\rangle)/\mu^2$ is the equivalent of Eq.~(\ref{eq:C}) for sources. Here the average total transverse momentum per source is defined as $\bar{P}_T=\int \hat{\rho}_1(\mathbf{p})\, p_t\, d\mathbf{p}$ and, using (\ref{eq:evnt_Ksrc_singles}), the average total transverse momentum for events is $\langle P_T\rangle = \langle K\rangle \bar{P}_T$. Following (\ref{eq:avgpt}) and substituting (\ref{eq:avgN_src}), the event averaged transverse momentum per particle is equivalently written as 
\begin{equation}
	\langle p_t\rangle = \bar{P}_T/ \mu.
\end{equation}
Finally, the variance of total transverse momenta per source is $\sigma^2_{P_T} =\overline{P^2_T} - \bar{P}_T^2$, where $\iint \hat{\rho}_2(\mathbf{p}_1, \mathbf{p}_2)\, p_{t,1}p_{t,2}\, d\mathbf{p}_1d\mathbf{p}_2 = \sigma^2_{P_T} + \mu^2\langle p_t\rangle^2-\mu\langle p_t\rangle$.\smallskip

Notice that both Eqs. (\ref{eq:R_ISM}) and (\ref{eq:C_ISM}) have similar contribution from the fluctuation in the number of sources. Given that the sources are independent, (\ref{eq:C_ISM}) decreases with the inverse of the number of sources in the same way as (\ref{eq:R_ISM}). However, momentum correlations (\ref{eq:C_ISM}) are sensitive to the transverse expansion due to the correlation function weighting by \pt. We reiterate that when \C~is measured as defined in  (\ref{eq:Cint}) and \textit{not} differentially in relative azimuthal angle or pseudorapidity, then effects form anisotropic flow are eliminated. \C~then represents the magnitude of transverse momentum correlations generated in the fireball. A measured deviation from predictions of the independent source model might suggest that sources of correlations are not independent, as would be the case for a partially or fully equilibrated system.

Multiplicity-momentum correlations, \D, are defined by (\ref{eq:Dint}). Following the same procedure we use for \R~and \C, we obtain
\begin{equation}\label{eq:D_ISM}
	{\cal D} = 
	\frac{{\cal D}_s}{\langle K\rangle}
\end{equation}
where ${\cal D}_s = \left( Cov(P_T,n) - \langle p_t \rangle \sigma^2 \right)/\mu^2$ is the equivalent of Eq.~(\ref{eq:D}) for sources rather than events. 

Notice that since (\ref{eq:D}) is constructed to remove the effects of multiplicity fluctuations, (\ref{eq:D_ISM}) does not have the same dependence on source fluctuations as \R~or \C. However, all three observables \R, \C, and \D~still are reduced by the inverse of the number of sources.

Lastly, correlations of transverse momentum fluctuations are defined by (\ref{eq:dptdptcorr}). Again, using Eq.~(\ref{eq:CorrFn}) with (\ref{eq:evnt_Ksrc_singles}) and (\ref{eq:evnt_Ksrc_pairs}) we find
\begin{eqnarray}
	\langle \delta p_{t1} \delta p_{t2}  \rangle &=& 
	\frac{\langle K\rangle\left({\cal C}_s-2\langle p_t\rangle{\cal D}_s-\langle p_t\rangle^2{\cal R}_s\right)}
	{\langle K\rangle{\cal R}_s + \langle K^2\rangle} \label{eq:dptdpt_ISM} \\
	\nonumber \\ 
	&=& \frac{\langle \delta p_{t1} \delta p_{t2}\rangle_s}{\langle K\rangle}
	\frac{(1+{\cal R}_s)}{(1+{\cal R})}. \label{eq:dptdpt_ISM2}
\end{eqnarray}
Here $(1+{\cal R}_s)\langle \delta p_{t1} \delta p_{t2} \rangle_s = {\cal C}_s-2\langle p_t\rangle{\cal D}_s-\langle p_t\rangle^2{\cal R}_s$ following the same reasoning leading to Eq. (\ref{eq:sumrule}) except that averaging is done over the ensemble of all possible independent sources rather than events. The denominator of (\ref{eq:dptdptcorr}) is different from the other observables in this work, but since \dptdpt~is well studied in literature, this form better suits direct comparison to measured data. The consequence is that the effects from fluctuating independent sources are not as obvious as the other observables. By examining (\ref{eq:dptdpt_ISM}) we see that \dptdpt~approximately decreases like $\langle K\rangle^{-1}$ in the limit of large $K$ where \R~is small. 

If we take the origin of the sources to be participant nucleons, then the minimum number of sources in any collision is two.  Calculations of ${\cal R}_s$, ${\cal C}_s$, ${\cal D}_s$, and $\langle \delta p_{t1} \delta p_{t2}\rangle_s$ in proton-proton collisions can serve as a possible representation of independent source correlations. In this scenario, \pp~collisions always have $K=2$ and never have a variance in the number of sources. Therefore we must have $\langle K^2\rangle - \langle K\rangle^2=0$ in (\ref{eq:R_ISM}) and (\ref{eq:C_ISM}). Taking (\ref{eq:R_ISM}) as an example for \pp~collisions, we have ${\cal R}_{pp} = {\cal R}_s/2$ for $K=2$ participants.  Consequentially for $AA$ collisions using $K=N_{part}$ and ${\cal R}_{s} = 2{\cal R}_{pp}$, we find Eq. (\ref{eq:R_WN}). Similarly,  Eqs.~(\ref{eq:R_ISM}), (\ref{eq:C_ISM}), (\ref{eq:D_ISM}), and (\ref{eq:dptdpt_ISM2}) become
\begin{eqnarray}
	{\cal R} &=& 
	\frac{2\,{\cal R}_{pp}}{\langle N_{part}\rangle} +
	\frac{\langle N_{part}^2\rangle - \langle N_{part}\rangle^2}{\langle N_{part}\rangle^2} \label{eq:R_WN} \\
	{\cal C} &=& 
	\frac{2\,{\cal C}_{pp}}{\langle N_{part}\rangle} +
	\left(\frac{\langle N_{part}^2\rangle - \langle N_{part}\rangle^2}{\langle N_{part}\rangle^2}\right)\langle p_t\rangle^2 \label{eq:C_WN} \\
	{\cal D} &=& \frac{2\,{\cal D}_{pp}}{\langle N_{part}\rangle} \label{eq:D_WN} \\
	\hspace{-9mm}\langle \delta p_{t1} &\delta p_{t2}&  \rangle = 
	\frac{2\,\langle \delta p_{t1} \delta p_{t2}\rangle_{pp}}{\langle N_{part}\rangle}
	\frac{(1+{\cal R}_{pp})}{(1+{\cal R})}. \label{eq:dptdpt_WN}
\end{eqnarray}
where \R~in the denominator of (\ref{eq:dptdpt_WN}) must come from (\ref{eq:R_WN}).

\bibliography{RDC_references}

\end{document}